\documentclass[preprint,review,12pt]{elsarticle} 
\usepackage{amssymb}
\usepackage{graphicx}
\usepackage{caption}
\usepackage{subcaption}
\usepackage{amsmath}
\usepackage{amssymb}
\usepackage{tikz}
\usepackage{secdot}
\usepackage{epstopdf}
\usepackage{geometry}
\usepackage{hyperref}
\usepackage{tabularx,ragged2e,booktabs,caption}
\usepackage{cite}
\usepackage{float}
\usepackage{lipsum}
\usepackage{fixltx2e}
\usepackage[normalem]{ulem}
\usepackage{multirow}
\graphicspath{ {./} }
\biboptions{numbers,sort&compress}
\journal{Journal of Solid State Chemistry}

\begin{document}
	
	\begin{frontmatter}
		
		\title{Crystalline phases in Zr$_9$Ni$_{11}$ and Hf$_9$Ni$_{11}$
			intermetallics; Investigations by perturbed angular correlation spectroscopy
			and ab initio calculations 
		}
		\author{S.K. Dey$^{1,2}$}
		\ead{skumar.dey@saha.ac.in}
		\author{C.C. Dey$^{1,2}$\corref{cor1}}
		\cortext[cor1]{corresponding author}
		\ead{chandicharan.dey@saha.ac.in}
		\author{S. Saha$^{1,2}$}
		\ead{satyajit.saha@saha.ac.in}
		\author{G. Bhattacharjee$^{1,2}$}
		\ead{gourab.bhattacharjee@saha.ac.in}
		\author{J. Belo$\check{\text{s}}$evi\'{c}-$\check{\text{C}}$avor$^3$}
		\ead{cjeca@vin.bg.ac.rs}
		\author{D. Toprek$^3$}
		\ead{toprek@vin.bg.ac.rs}
		\address{$^1$Saha Institute of Nuclear Physics, 1/AF Bidhannagar, Kolkata-700 064, India}
		\address{$^2$ Homi Bhabha National Institute, Anushaktinagar, Mumbai-400 094, India}
		\address{$^3$Institute of Nuclear Sciences Vinca, University of Belgrade, P. O. Box 522,\\ 11001 Belgrade, Serbia}
\begin{abstract}
Crystalline phases formed in stoichiometric Zr$_9$Ni$_{11}$ and Hf$_9$Ni$_{11}$ have been studied by perturbed angular correlation (PAC) spectroscopy, XRD and TEM/SAED measurements. In Zr$_9$Ni$_{11}$, the phases Zr$_9$Ni$_{11}$ ($\sim$89\%) and Zr$_8$Ni$_{21}$ ($\sim$11\%) have been found at room temperature from PAC measurements. At 773 K, Zr$_9$Ni$_{11}$ partially decomposes to Zr$_7$Ni$_{10}$ and at 973 K, it is completely decomposed to ZrNi and Zr$_7$Ni$_{10}$. In Hf$_9$Ni$_{11}$, a predominant phase ($\sim$81\%) due to 
HfNi is found at room temperature while the phase Hf$_9$Ni$_{11}$ is produced as a minor phase ($\sim$19\%). No compositional phase
change at higher temperature is found in Hf$_9$Ni$_{11}$. Phase components found from XRD and TEM/SAED measurements are similar to those
observed from PAC measurements. Electric field gradients in Zr$_9$Ni$_{11}$ and Hf$_9$Ni$_{11}$ have been calculated by density functional theory (DFT) using all electron full potential (linearized) augmented plane wave plus local orbitals [FP-(L)APW+lo] method in order to assign
the phase components.
\end{abstract}

\end{frontmatter}

\section{Introduction}
The Zr-Ni intermetallic compounds are widely used to synthesize advanced materials \citep{Pang,Ivanchenko,Haour,Koch}. The compounds Zr$_8$Ni$_{21}$, Zr$_7$Ni$_{10}$,
Zr$_9$Ni$_{11}$ and ZrNi are, particularly, important due to their potential application as gaseous hydrogen storage materials. The electrochemical properties
of these compounds were extensively
studied earlier \citep{Ruiz,FCRuiz,Joubert,Nei}. 
These alloys absorb gaseous hydrogen in their interstitial sites to form nickel metal hydrides (NiMH). These metal hydrides are used as
negative electrode in NiMH rechargeable
batteries. It was found that the hydrogen absorption/desorption properties of many Laves phase and quaternary/ternary hydriding alloys are improved due to
the presence of non-Laves Zr-Ni binary alloys \citep{Zhang, Young, Ouchi, Kwo, Fetcenko, Akihiro}.
The NiMH batteries have high energy density compared to nickel-cadmium batteries and are more environmentally friendly. 
Recently, hydrogen absorption/desorption properties in ZrNi alloy were investigated \citep{Akihiro} by partial substitution of Zr with Ti for
different Ti concentration. 
Also, in case of Zr$_9$Ni$_{11}$, unusual magnetic properties were
reported by Provenzano et al. \citep{ProvenzanoZr9Ni11}. Its crystal structure is known to be body-centered tetragonal
with space group $I4/m$ \citep{Glimois,KrikpatrickLarson,Shadangi,Xueyan,Akihiro,Joubert,Kosorukova,Nash} which is reported to be
isostructural to Zr$_9$Pt$_{11}$ \citep{Panda,Shadangi,Bhan}. 
The lattice parameters of Zr$_9$Ni$_{11}$ were found to be $a$ = 9.88(1) $\AA$ and $c$ = 6.61(1) $\AA$ \citep{Glimois}. The Hf$_9$Ni$_{11}$ was found
to be isostructural to Zr$_9$Ni$_{11}$, with the lattice parameters $a$ = 9.79 $\AA$ and $c$ = 6.53 $\AA$ \citep{KrikpatrickLarson, Liu, PAnash}. 

The phase diagram of Zr-Ni system \citep{KrikpatrickLarson,Nash,Ghosh,Kosorukova,Bsenko} shows at least eight binary compounds. From a recent study \citep{Xiaoma}, all these phases were found through the interdiffusion process by annealing the Zr-Ni interface zone at different temperatures. It is found that the phases
Zr$_2$Ni$_7$, Zr$_2$Ni, and ZrNi melt congruently while Zr$_8$Ni$_{21}$, Zr$_9$Ni$_{11}$, Zr$_7$Ni$_{10}$, and ZrNi$_5$ phases form 
peritectically. The phase ZrNi$_3$ forms by peritectoid reaction. The room temperature formation of these phases are due to nucleation by rapid quenching
from the liquid
alloy \citep{Ghosh, Akihiro}.
Stalick et al. \citep{StalickZr9Ni11} found that the Zr$_9$Ni$_{11}$ sample, prepared by argon arc melting, contained 10\% of ZrNi and at 700 $^\circ$C, it 
transformed to ZrNi (37\%), Zr$_7$Ni$_{10}$ (57\%) and Zr$_9$Ni$_{11}$ (7\%). At 1000 $^\circ$C, this sample reformed again to produce
90\% Zr$_9$Ni$_{11}$ and 10\% ZrNi.
The compounds with Zr or Hf as a constituent element are very suitable for perturbed angular correlation (PAC) measurements
using the probe $^{181}$Hf. Due to chemical similarity, the probe $^{181}$Hf occupies the Zr sites of the compound.
Different Zr-Ni and Hf-Ni intermetallic
compounds viz., (Zr/Hf)$_8$Ni$_{21}$ \citep{skdeyZr8Ni21}, (Zr/Hf)$_2$Ni$_{7}$ \citep{CCDeyPhysica,Marszalek}, (Zr/Hf)Ni$_{3}$ \citep{SKDey_JACOM} were 
studied earlier by PAC technique. However, any report of
PAC measurements in the (Zr/Hf)$_9$Ni$_{11}$ systems are not available in literature.
The technological applications of Zr$_9$Ni$_{11}$ and absence of PAC measurements,
promoted the study of Zr$_9$Ni$_{11}$ and Hf$_9$Ni$_{11}$ compounds to determine the electric field
gradients (EFG) at the Ta impurity sites, obtained after radioactive decay of $^{181}$Hf and also the structural stability of the compounds.
The measured EFG and asymmetry parameters can be directly compared to the calculated
values obtained from first-principles density functional theory (DFT), which helps 
assigning the component phases produced in the sample.

The time-diffrential perturbed angular correlation (TDPAC), also called PAC, is a nuclear technique based on the interaction of electromagnetic moments
(electric quadrupole or magnetic dipole) of a specific nuclear level
with the electric field gradient or the magnetic field generated at
the nuclear level by the surrounding environment of the nucleus. Electric field gradient strongly depends upon the lattice
parameters and the crystal structure of the lattice. This technique is sensitive to the change of local field environment
of the probe nucleus. Production of multiphase components can be identified by this technique
from the observation of different quadrupole frequencies. In the present report, phase stability and structural phase transformation in Zr$_9$Ni$_{11}$ and 
Hf$_9$Ni$_{11}$ have been studied by observing the temperature dependence of PAC parameters. Additionally, we have carried out X-ray diffraction (XRD) and 
transmission electron microscopy (TEM)/selected area electron diffraction (SAED) measurements to determine the phase components in these samples. Calculations of EFG and 
asymmetry parameter ($\eta$) at $^{181}$Ta impurity site have also been carried out in Zr$_9$Ni$_{11}$ and Hf$_9$Ni$_{11}$ to compare with the experimental results from PAC measurements and therefrom to assign these phases. 

\section{Experimental details}

The intermetallic compounds Zr$_9$Ni$_{11}$ and Hf$_9$Ni$_{11}$ have been prepared in argon arc furnace using the stoichiometric amounts of constituent elements
procured from Alfa Aesar. 
The purity of Zr (excluding Hf), Hf (excluding Zr) and Ni were 99.2\%, 99.95\% and 99.98\%, respectively. The probe $^{181}$Hf 
was produced in Dhruba reactor, Mumbai by irradiating natural Hf metal {($\sim$30\% $^{180}$Hf)} with thermal neutron for 7 days (flux $\sim$10$^{13}$/cm$^2$/s). For PAC
measurements, this active Hf metal was added {with $\sim$0.3 at\%} to the compounds and remelted in the arc furnace. {It can be considered that addition of this very small impurity concentration will not affect stoichiometry of the samples and the sample properties.} Separate inactive
samples were produced in similar manners for structural characterization by XRD and TEM/SAED measurements. Barring in Hf$_9$Ni$_{11}$ PAC sample, no
appreciable mass loss of other samples were found during arc melting. In all cases, shiny globule samples were formed. The X-ray diffraction measurements
have been carried out using Rigaku X-ray diffractometer TTRAX-III and Cu K$_\alpha$ radiation. Transmission electron
microscopy (TEM) measurements were carried out using
FEI, Tecnai G2 F30, S-Twin microscope equipped with a high angle annular dark-field (HAADF) detector, a
scanning unit and a energy dispersive X-ray spectroscopy (EDX) unit to
perform the scanning transmission electron
microscopy (STEM-HAADF-EDX).

The $^{181}$Hf probe is introduced into the investigated sample (Zr/Hf)$_9$Ni$_{11}$  which replaces Zr/Hf atoms in the matrix. This probe atom {$^{181}$Hf $\beta^-$ decays to $^{181}$Ta ($T_{1/2}$ = 42.4 d) and populates the 615 keV excited level of $^{181}$Ta. The daughter nucleus $^{181}$Ta comes to ground state through the successive gamma rays of 133 and 482 keV passing through the 482 keV intermediate level.} The intermediate level has
a half-life of 10.8 ns and a spin angular momentum $I$ = 5/2$^+\hbar$ \citep{Firestone}. {Since the $\beta^-$ decay half life of $^{181}$Hf is 42.4 d, the PAC measurements are generally performed within 1-2 months of probe activation i.e. before the probe is decayed to a very low activity level.} The extra-nuclear electric field gradient (EFG) present in the sample interacts with
the quadrupole moment of the probe nucleus and perturbs
the $\gamma$-$\gamma$ angular correlation. {Thus, through the PAC technique, the EFG in a sample is actually  measured at the $^{181}$Ta impurity atom.}

The perturbation function in a polycrystalline sample for I=5/2 $\hbar$ is given by \citep{Schatz, Zacate}
\begin{equation}
G_2(t)=S_{20}(\eta) + \sum^{3}_{i=1}S_{2i}(\eta)\text{cos}(\omega_it)\text{exp}(-\delta\omega_it).
\label{eqn:Stokes}
\end{equation}
The
frequencies $\omega_i$ denote the transition frequencies between the $m$-sublevels of the intermediate state. These sublevels become
non-degenerate in energies due to hyperfine splitting. The damping in PAC spectrum is represented by $\delta$ (Lorentzian distribution) which arises due to lattice imperfections
and chemical inhomogeneities of the sample. The three $\omega_i$'s in the
perturbation function $G_2(t)$ are related to $V_{zz}$, the maximum component of EFG in the principle axis system, through the quadrupole frequency given by 
\begin{equation}
\omega_Q= \frac{eQV_{zz}}{4I(2I-1)\hbar}.
\label{eqn:raman}
\end{equation}
In the above Eqn. \ref{eqn:raman}, $Q$ represents the quadrupole moment of the intermediate level (2.36 b \citep{Butz}). 
For an axially symmetric EFG ($\eta=0$), the $\omega_Q$ is related to $\omega_1$, $\omega_2$ and $\omega_3$ by 
\begin{equation}
\omega_Q=\omega_1/6=\omega_2/12=\omega_3/18.
\label{eqn:prafulla}
\end{equation}
The principal EFG components obey the relations
\begin{equation}
V_{xx} + V_{yy} + V_{zz}=0 \quad \text{and}\quad
|V_{zz}|\ge |V_{yy}|\ge |V_{xx}|.
\label{eqn:hizenberg}
\end{equation}
The EFG can therefore be designated by two parameters only viz. $V_{zz}$ and $\eta$. The asymmetry parameter $\eta$ is defined as  
\begin{equation}
\eta=\frac{(V_{xx}-V_{yy})}{V_{zz}},\quad \text{}\quad 0\le\eta\le1.
\label{eqn:newton}
\end{equation} 
For $\eta\ne$0, a more general relation between $\omega_Q$ and $\omega_i$'s is found \citep{Zacate}.

A slow-fast coincidence set up comprising two LaBr$_3$(Ce) and two BaF$_2$ detectors has been built to acquire data simultaneously at 180$^\circ$ and 90$^\circ$ in a
coplanar arrangement. Crystal sizes were
38$\times$25.4 mm$^2$ for LaBr$_3$(Ce) and 50.8$\times$50.8 mm$^2$ for BaF$_2$ scintillators. The low energy $\gamma$-rays (133 keV) of the 
probe nucleus were captured in LaBr$_3$(Ce) detectors. A typical prompt time resolution of $\sim$800 ps has been observed for the
LaBr$_3$(Ce)-BaF$_2$ detector combination selecting the 133-482 keV $\gamma$-rays of $^{181}$Hf. The perturbation function $G_2(t)$ is obtained from the four coincidence spectra taken at 180$^\circ$ and 90$^\circ$ \citep{pramana}.

\begin{table}
	\begin{center}
\caption{\label{tab:Zr9Ni11_table}Results of PAC measurements in Zr$_9$Ni$_{11}$}
\scalebox{0.68}{
\begin{tabular}{ccccccc}
\hline
Temperature (K)  &Component    & $\omega_Q$ (Mrad/s)     & $\eta$     & $\delta$($\%$)   & $f$($\%$)      & Assignment   \\ 
\hline  	
				
77             &1        & 41(1)                 & 0.80(10)          & 17(3)            & 84(3)      & Zr$_9$Ni$_{11}$        \\   
				&2      & 76.8(8)                 & 0.53(3)          & 0                & 16(3)         &    Zr$_8$Ni$_{21}$\\ \\  
				
298             &1        & 39.2(5)                 & 0.64(3)          & 15(2)            & 89(3)      &   Zr$_9$Ni$_{11}$      \\   
				&2      & 77.5(7)                 & 0.52(2)          & 0                & 11(3)         &    Zr$_8$Ni$_{21}$ \\   \\
				
373               &1     & 39.0(8)                 & 0.74(5)          & 12(3)                & 80(3)    & Zr$_9$Ni$_{11}$           \\   
				&2      & 76.8(7)                 & 0.51(2)          & 0                & 20(3)  &        Zr$_8$Ni$_{21}$  \\    \\
				
473              &1      & 39.4(8)                 & 0.59(3)          & 17(3)                & 85(3)    &  Zr$_9$Ni$_{11}$           \\ 
				&2      & 75.5(7)                 & 0.58(3)          & 0                & 15(3)  &        Zr$_8$Ni$_{21}$  \\    \\
				
573              &1      & 39.9(5)                 & 0.58(2)          & 12(2)                & 82(3)   &   Zr$_9$Ni$_{11}$         \\   
				&2      & 75.9(5)                 & 0.57(2)          & 0                & 18(3)  &        Zr$_8$Ni$_{21}$  \\    \\
				
673              &1      & 37.5(5)              & 0.57(3)          & 12(2)               & 81(3)  &      Zr$_9$Ni$_{11}$         \\  
				&2      & 75.3(6)                 & 0.51(2)          & 0                & 19(3)  &      Zr$_8$Ni$_{21}$  \\ \\
				
773              &1      & 38.4(4)                 & 0.48(3)        & 2(1)                & 45(3)  &     Zr$_9$Ni$_{11}$     \\  
				&2      & 73.3(8)                 & 0.61(3)          & 0                & 20(3)  &        Zr$_8$Ni$_{21}$  \\    
				&3      & 45.8(8)                & 0.57(6)                & 0                & 27(3)   &  Zr$_7$Ni$_{10}$           \\
				&3      & 99(2)                & 0                & 0                & 7(3)   &           \\  \\
				
873              &1      & 38.5(7)                 & 0.38(4)          & 8(4)                & 54(3)    &    Zr$_9$Ni$_{11}$   \\   
				&2      & 76.8(8)                 & 0.66(2)          & 0                & 18(3)  &      Zr$_8$Ni$_{21}$     \\  
				&3      & 46.0(9)                 & 0.60(8)               & 0                & 16(3)  &  Zr$_7$Ni$_{10}$       \\    
				&4      & 96(1)                & 0        & 0                & 11(3)  &            \\ \\        
				
973              &1      & 17.9(6)                 & 0.36(4)          & 0                & 55(3)      &  ZrNi   \\
				&2      & 38.2(4)                 & 0.52(5)          & 0                & 32(3)  &        Zr$_9$Ni$_{11}$  \\    
				&3      & 45(1)                & 0.72(12)          & 0                & 13(3)      &  Zr$_7$Ni$_{10}$   \\       \\ 
				
298$^a$       &1    & 26.2(6)                 & 0.40(3)         & 8(2)                & 63(3)         &  ZrNi\\ 
&2   & 74.2(4)                & 0          & 0                & 17(3)    &  ZrNi$_3$   \\   
&3   & 77(1)                & 0.68(4)          & 0                & 11(3)    &   Zr$_8$Ni$_{21}$\\
&4   & 62(1)                & 0.60(8)          & 0                & 9(3)    &   Zr$_7$Ni$_{10}$ \\  
\hline                                             
\end{tabular}}
\end{center}
\begin{flushleft}
	{$^a$ after measurement at 973 K}
\end{flushleft}

\end{table}

	\begin{figure*}[t!]
		\centering
		\begin{subfigure}[t]{0.40\textwidth}
			\centering
			\includegraphics[scale=.38]{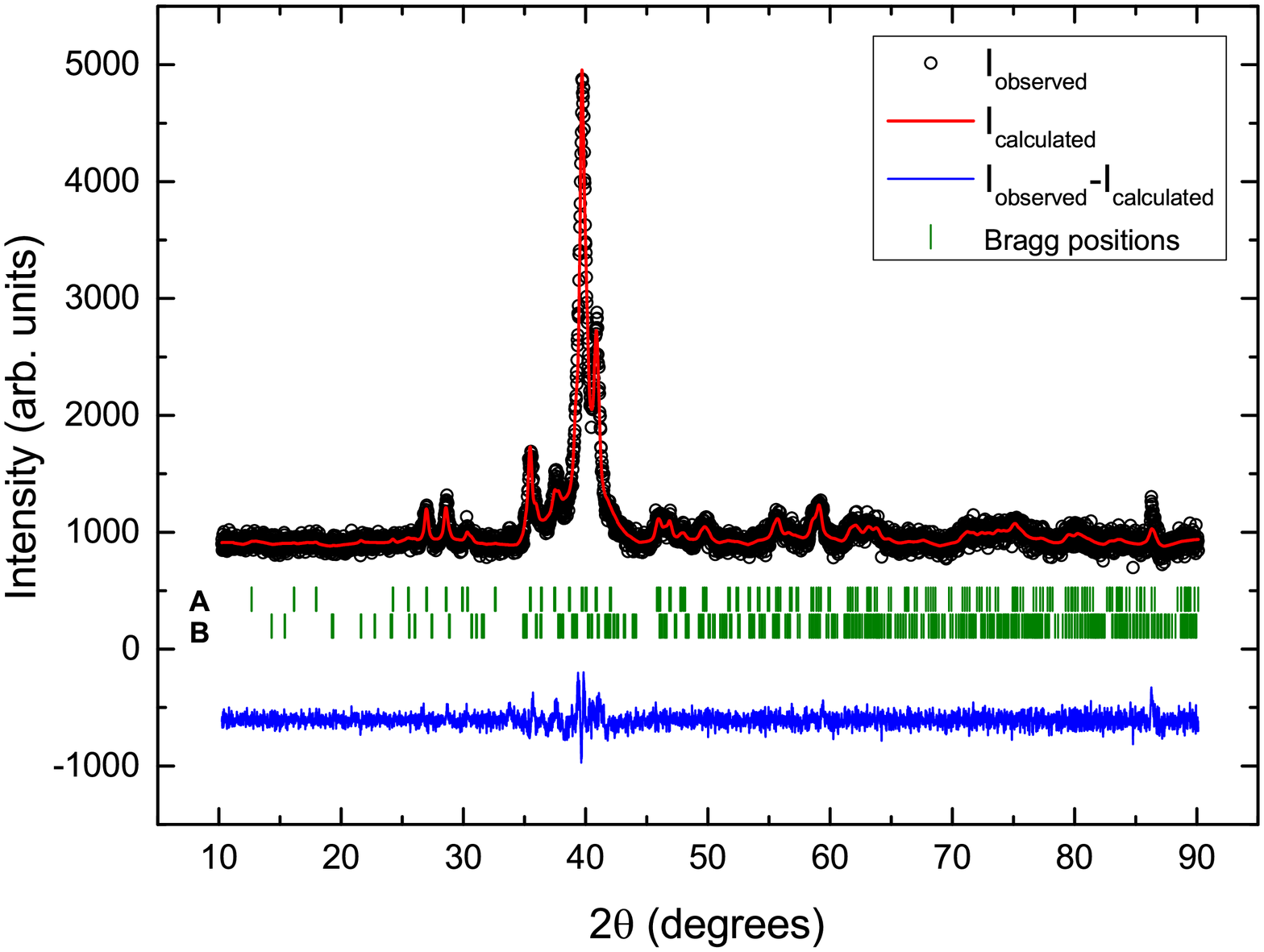}
		\end{subfigure}
		\hspace{2.7cm}
		\begin{subfigure}[t]{0.4\textwidth}
			\centering
			\includegraphics[scale=.18]{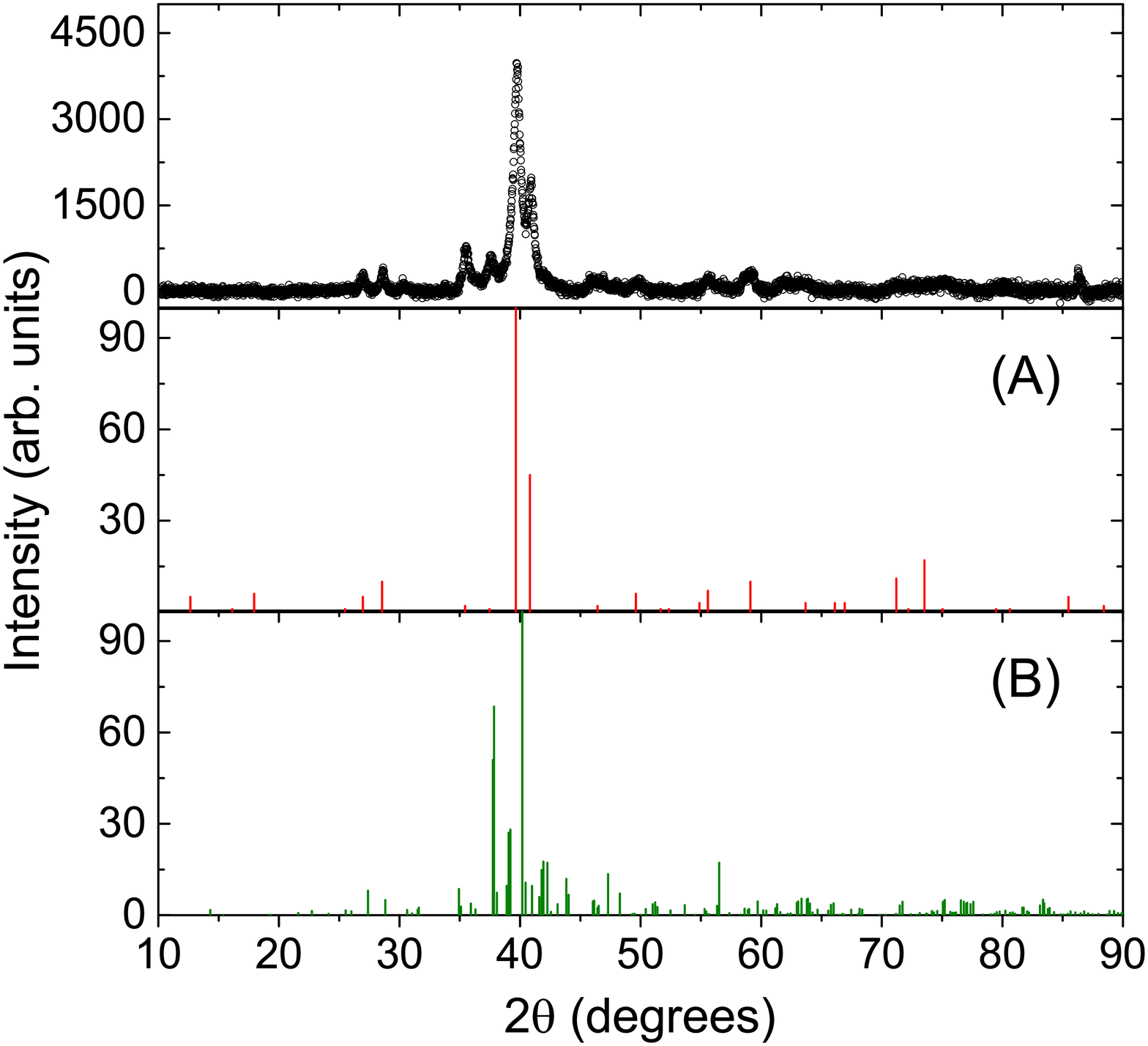}
		\end{subfigure}
		\caption{Figure in the left shows the background subtracted XRD powder pattern in the stoichiometric sample of Zr$_9$Ni$_{11}$. The line represents the fit to the measured data. The vertical bars A, B denote the Bragg angles corresponding to Zr$_9$Ni$_{11}$ and Zr$_7$Ni$_{10}$, respectively. The bottom line shows the difference between the observed and the fitted pattern. {Figure in the right shows a comparison of ICDD pattern of Zr$_9$Ni$_{11}$ (A), Zr$_7$Ni$_{10}$ (B) with the experimental XRD pattern of Zr$_9$Ni$_{11}$}}
			\label{fig:Zr9Ni11ascast_XRD}
	\end{figure*}

%
%
%
%
%
%
%
\begin{figure}[t!]
\begin{center}
\includegraphics[width=0.6\textwidth]{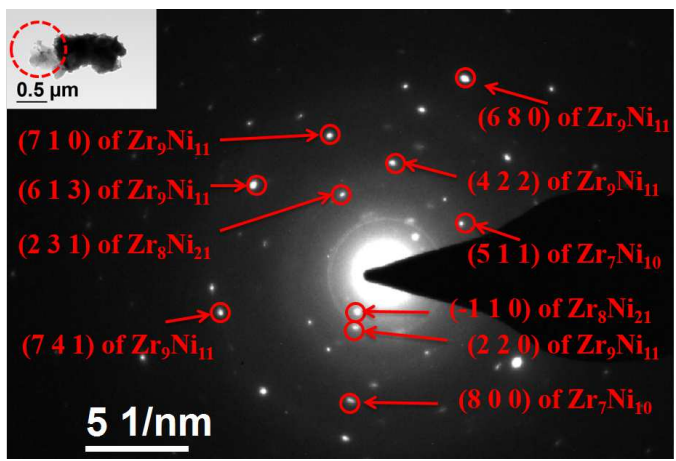}
\end{center}
\caption{\label{fig:Zr9Ni11ascast_TEM}Selected area electron diffraction pattern from the stoichiometric Zr$_9$Ni$_{11}$ particle shown in the inset.}
\end{figure}
		
\begin{figure*}[t!]
\begin{center}
\includegraphics[width=0.7\textwidth]{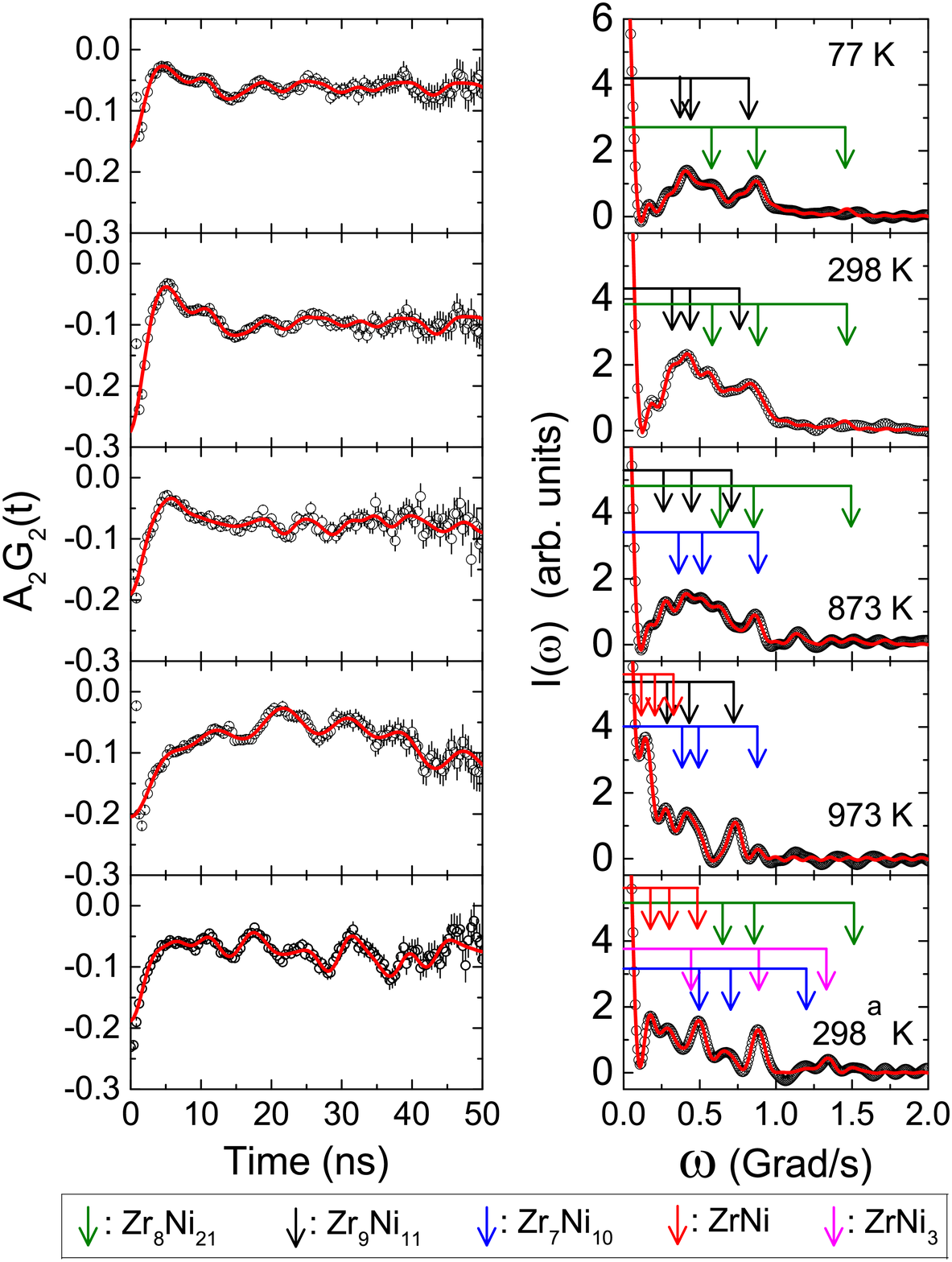}
\end{center}
\caption{\label{fig:Zr9Ni11_spectra}PAC spectra in the stoichiometric sample of Zr$_9$Ni$_{11}$ at different temperature. Left panel shows the time spectra and the right panel shows the corresponding Fourier cosine transforms. The PAC spectrum at room temperature designated by 298$^a$ K is taken after the measurement at 973 K.}
\end{figure*} 
		
\begin{figure}[h!]
\begin{center}
\includegraphics[width=0.45\textwidth]{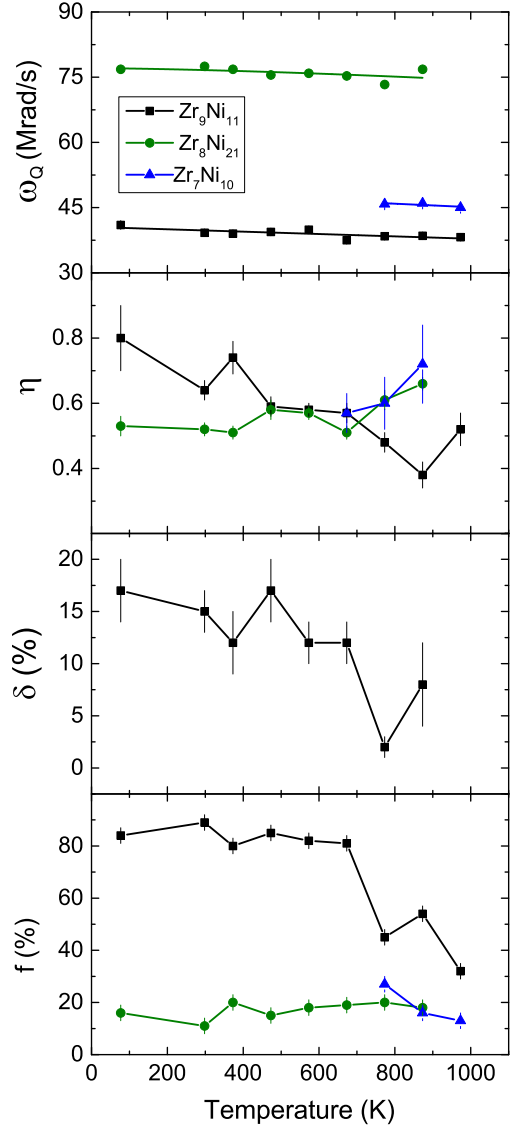}
\end{center}
\caption{\label{fig:Zr9Ni11_parameter} Variations of quadrupole frequency ($\omega_Q$), asymmetry parameter ($\eta$), frequency distribution width ($\delta$) and site fraction $f$(\%) with temperature for the components of Zr$_9$Ni$_{11}$, Zr$_8$Ni$_{21}$ and Zr$_7$Ni$_{10}$.}		
\end{figure}

	\begin{figure*}[t!]
		\centering
		\begin{subfigure}[t]{0.40\textwidth}
			\centering
			\includegraphics[scale=.38]{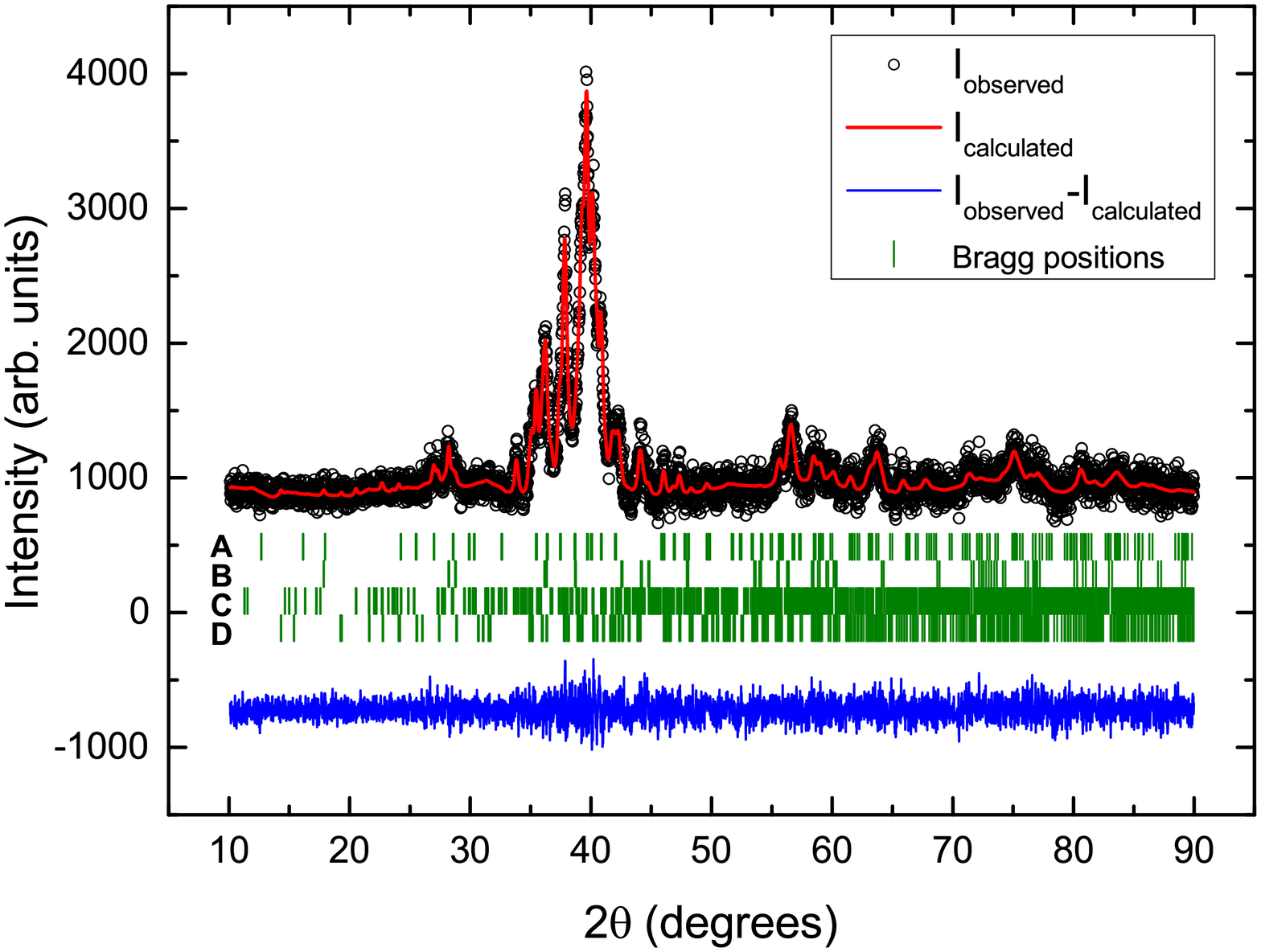}
		\end{subfigure}
		\hspace{2.7cm}
		\begin{subfigure}[t]{0.4\textwidth}
			\centering
			\includegraphics[scale=.18]{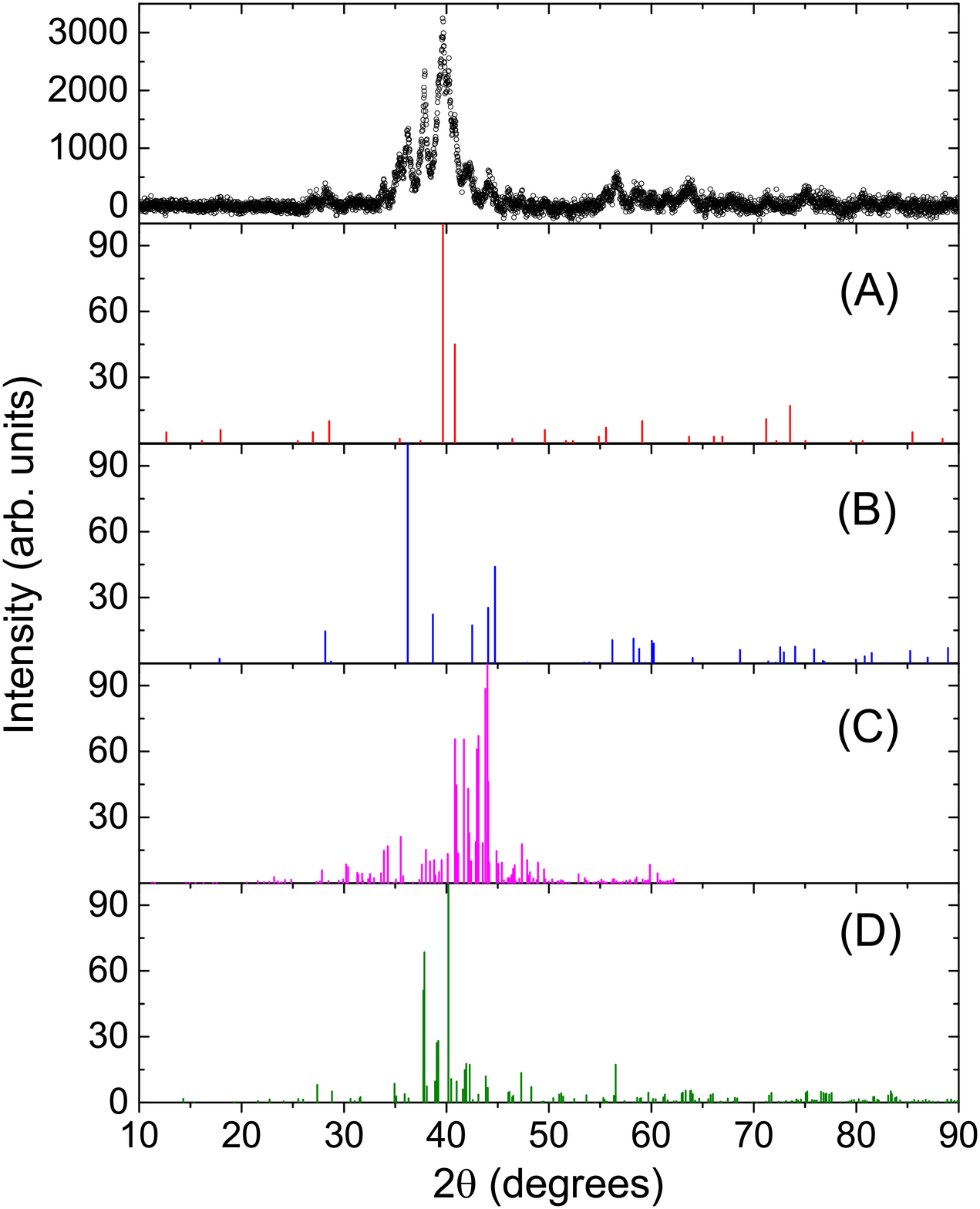}
		\end{subfigure}
		\caption{Figure in the left shows the background subtracted XRD powder pattern in the stoichiometric
			sample of Zr$_9$Ni$_{11}$ after the sample is annealed at 800$^\circ$C for two days. The line represents the fit to the measured data. The vertical bars A, B, C and D denote the Bragg angles corresponding to Zr$_9$Ni$_{11}$, ZrNi, Zr$_8$Ni$_{21}$ and Zr$_7$Ni$_{10}$, respectively. The bottom line shows the difference between the observed and the fitted pattern. {Figure in the right shows a comparison of ICDD pattern of Zr$_9$Ni$_{11}$ (A), ZrNi (B), Zr$_8$Ni$_{21}$ (C) and Zr$_7$Ni$_{10}$ (D) with the experimental XRD pattern of the annealed Zr$_9$Ni$_{11}$ sample.}}
		\label{fig:Zr9Ni11annealed_XRD}
	\end{figure*}
%
%
%
%
%

\begin{figure}[t!]
\begin{center}
\includegraphics[width=0.6\textwidth]{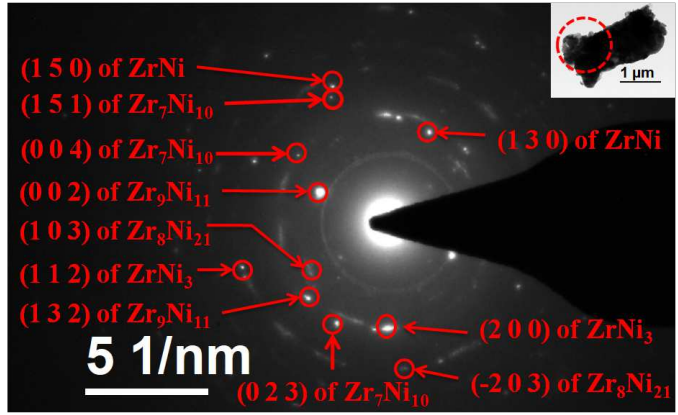}
\end{center}
\caption{\label{fig:Zr9Ni11annealed_TEM}Selected area electron diffraction pattern from stoichiometric Zr$_9$Ni$_{11}$ particle (after annealing at 1073 K for 2 days) shown in the inset.}
\end{figure}
	
\section{Results and discussion}
\subsection{Zr$_9$Ni$_{11}$} 
		The XRD powder pattern in Zr$_9$Ni$_{11}$ is shown in Fig.~\ref{fig:Zr9Ni11ascast_XRD}. {The peaks were identified (Fig. \ref{fig:Zr9Ni11ascast_XRD}) using ICDD database, 2009.} Presence
		of tetragonal Zr$_9$Ni$_{11}$ [PDF Card No.: 00-033-0963] and orthorhombic Zr$_7$Ni$_{10}$ [PDF Card No.: 01-072-3501] phases have been found in the XRD spectrum. Analysis of
		the X-ray powder pattern (Fig. \ref{fig:Zr9Ni11ascast_XRD}) was
		carried out by FullProf software package \citep{Rodriguez} using the known crystallographic data 
		of Zr$_9$Ni$_{11}$ with space group $I4/m$ \citep{Glimois} and Zr$_7$Ni$_{10}$ \citep{Yvon}. The presence of a minor phase Zr$_7$Ni$_{10}$ in the stoichiometric sample of Zr$_9$Ni$_{11}$ 
		was also observed from x-ray diffraction measurements by Joubert et al. \citep{Joubert}.
		
		The presence of Zr$_9$Ni$_{11}$, Zr$_8$Ni$_{21}$ and Zr$_7$Ni$_{10}$ phases in
		this stoichiometric sample of Zr$_9$Ni$_{11}$ have been observed from TEM/SAED measurement (Fig.~\ref{fig:Zr9Ni11ascast_TEM}). The 
		SAED pattern obtained from a region marked by a dotted circle in the stoichiometric sample of Zr$_9$Ni$_{11}$ is shown in Fig.~\ref{fig:Zr9Ni11ascast_TEM}. The interplanar spacing ($d_{hkl}$) is obtained by
		measuring the distance ($\Delta q$) of a particular spot from the central bright spot using the formula $d_{hkl}=1/\Delta q$. Some of the 
		measured $d_{hkl}$ from the SAED pattern
		are 1.39 \r{A}, 1.30 \r{A}, 1.21 \r{A}, 3.45 \r{A}, 0.99 \r{A} and 1.84 \r{A}. These measured interplanar spacings are very close to the (710), (613), (741), (220), (680) and (422) interplanar
		spacings of tetragonal Zr$_9$Ni$_{11}$ (JCPDS $\#$ 33-0963), respectively. This further confirms the presence of Zr$_9$Ni$_{11}$ phase in the sample. 
		Few of the measured $d_{hkl}$ from the SAED pattern
		are  2.29 \r{A} and 4.34 \r{A} which are very close to the (231) and (-110) interplanar
		spacings of triclinic Zr$_8$Ni$_{21}$ (ICDD PDF Card No.: 01-071-2622), respectively. The presence of Zr$_8$Ni$_{21}$ phase in the sample is thus confirmed. 
		Some of the measured interplanar
		spacings from the SAED pattern, 2.32 \r{A} and 1.55 \r{A}, are found to be very close to the (511) and (800) interplanar
		spacings of orthorhombic Zr$_7$Ni$_{10}$ (ICDD PDF card No.:01-072-3501), respectively, which confirms the presence of Zr$_7$Ni$_{10}$ phase in the stoichiometric sample
		of Zr$_9$Ni$_{11}$.
		
		The PAC spectrum found at room temperature in the stoichiometric sample of Zr$_9$Ni$_{11}$ is shown in Fig.~\ref{fig:Zr9Ni11_spectra}. 
		The predominant component (89\%) produces values of $\omega_Q$ = 39.2(5) Mrad/s ($V_{zz}$ = 4.4(1)$\times$10$^{21}$ V/m$^2$), $\eta$ = 0.64(3) with a large frequency distribution
		width ($\delta$ = 15\%). Apart from this, a minor component is also observed (Table \ref{tab:Zr9Ni11_table}). The predominant component has been 
		assigned to Zr$_9$Ni$_{11}$ by comparing the results with DFT calculations (discussed later). The minor component found can be attributed to Zr$_8$Ni$_{21}$ by comparing the values of $\omega_Q$ and $\eta$ with the
		results from our recent investigation \citep{skdeyZr8Ni21} for a particular crystallographic site in Zr$_8$Ni$_{21}$. It was found \citep{Nash} that the phases Zr$_2$Ni$_7$, Zr$_2$Ni and ZrNi melt congruently. The phase Zr$_8$Ni$_{21}$ was
		formed at $\sim$1453 K from the liquid by peritectic reaction L+Zr$_2$Ni$_7\rightarrow$Zr$_8$Ni$_{21}$ \citep{Nash}. Kirkpatrick and
		Larson \citep{KrikpatrickLarson} reported that the phase Zr$_9$Ni$_{11}$ was produced peritectically by the reaction 
		L+ZrNi$\rightarrow$Zr$_9$Ni$_{11}$ at 1443 K \citep{Nash} and this phase
		was found to decompose eutectoidally Zr$_9$Ni$_{11}\rightarrow$ ZrNi+Zr$_7$Ni$_{10}$ at $\sim$1248 K \citep{Nash}. In the process of arc-melting, the
		production of Zr$_9$Ni$_{11}$ phase at
		room temperature is due to rapid solidification from the liquid melt.   
		To find any structural change in this compound or its phase stability with temperature, PAC measurements have been performed in
		the sample at high temperatures also (up to 973 K). The results
		show (Table \ref{tab:Zr9Ni11_table}) that up to 673 K, the phase due to Zr$_9$Ni$_{11}$
		remains stable and the component fractions do not change appreciably.
		At 773 K, however, the sample compositions change. At this temperature,
		a new component ($\sim$27\%) appears at the expense of Zr$_9$Ni$_{11}$. This has been attributed
		to Zr$_7$Ni$_{10}$ by comparing with the previous PAC results in Zr$_7$Ni$_{10}$ \citep{skdeyZr7Ni10}. Kosorukova et al. \citep{Kosorukova} reported
		that the phase Zr$_7$Ni$_{10}$ was produced by peritectic reaction L+Zr$_9$Ni$_{11}\rightarrow$ Zr$_7$Ni$_{10}$ at 1393 K. 
		The component fraction due to Zr$_8$Ni$_{21}$ at 773 K, however, does not change. Another new component with a small
		fraction (Table \ref{tab:Zr9Ni11_table}) also appears which could not be assigned. At 873 K, the phase due to Zr$_9$Ni$_{11}$ slightly
		increases to 54\% at the cost of Zr$_7$Ni$_{10}$ component. At
		a temperature of 973 K, the sample composition changes drastically. At this temperature,
		the predominant component ($\sim$55\%) is found to have $\omega_Q$ = 17.9(6) Mrad/s, $\eta$ = 0.36(4),
		$\delta$ = 0 and this can be attributed to ZrNi by comparing the results found in ZrNi
		from previous PAC measurements \citep{CCDeyJPCS}. 
		At this temperature, the Zr$_8$Ni$_{21}$ phase vanishes and the Zr$_9$Ni$_{11}$
		phase reduces to 32\%. It was reported \citep{KrikpatrickLarson,Nash} that the the phase ZrNi was formed by the decomposition of 
		Zr$_9$Ni$_{11}$ phase eutectoidally at $\sim$1248 K. We have carried out a re-measurement
		in the sample at room temperature after the measurement at 973 K to find the change
		in sample composition before and after heating the sample.
		In the re-measurement, a new component ($\sim$17\%) with
		$\omega_Q$ = 74.2(4) Mrad/s, $\eta$ = 0 is found. This new component can be attributed to ZrNi$_3$ by comparing the results reported from our recent
		PAC investigations in ZrNi$_3$ \citep{SKDey_JACOM}. It is known that ZrNi$_3$ was formed by the peritectoid reaction (Zr$_2$Ni$_7$+Zr$_8$Ni$_{21}\rightarrow$ZrNi$_3$) at 1193 K \citep{Nash} and at 1261 K \citep{Kosorukova}.
		The predominant phase
		due to ZrNi (63\%) is found to be produced here which was first observed at 973 K in the sample. Appearance of ZrNi as a secondary phase
		in the synthesized Zr$_9$Ni$_{11}$ alloy was also found earlier \citep{StalickZr9Ni11,Joubert,Xueyan,Zhang}.
		The remaining phases in the sample can be identified as Zr$_8$Ni$_{21}$ ($\sim$11\%) and Zr$_7$Ni$_{10}$ ($\sim$9\%). Interestingly, no phase due to Zr$_9$Ni$_{11}$ is found here and
		the sample is transformed completely. 
		
		The Zr$_9$Ni$_{11}$ compound was studied by Stalick et al. \citep{StalickZr9Ni11} earlier using transmission
		electron microscopy and powder neutron diffraction in the
		temperature range 4-1273 K. At room temperature,
		these authors reported a mass fraction of 90\% Zr$_9$Ni$_{11}$ and
		10\% ZrNi in Zr$_9$Ni$_{11}$. From present PAC measurements, however,
		no ZrNi phase is found at lower temperatures (below 873 K).
		The ZrNi phase is found only at 973 K. Stalick et al. \citep{StalickZr9Ni11} reported
		a sharp enhancement of this phase at 973 K and the production
		of a new phase of Zr$_7$Ni$_{10}$. The phase fractions were reported to be 37\% ZrNi, 57\% Zr$_7$Ni$_{10}$ and 7\% Zr$_9$Ni$_{11}$ at 973 K \citep{StalickZr9Ni11}. 
		We have also found phase components
		of ZrNi, Zr$_7$Ni$_{10}$ and Zr$_9$Ni$_{11}$ at 973 K although the phase fractions
		are found to be different (Table \ref{tab:Zr9Ni11_table}). An agreement
		between the results of Stalick et al. \citep{StalickZr9Ni11} and present
		results at high temperature are thus observed. Stalick et al. \citep{StalickZr9Ni11} found
		also a phase reformation at 1273 K where the sample
		was again 90\% Zr$_9$Ni$_{11}$ and 10\% ZrNi. Production of Zr$_7$Ni$_{10}$ and ZrNi can also be explained from the Zr-Ni phase diagram due to their proximity
		to Zr$_9$Ni$_{11}$ phase and very small homogeneity range of this phase \citep{Glimois, KrikpatrickLarson,Nash,Joubert}.
		
		The variations of $\omega_Q$, $\eta$, $\delta$ and phase fraction ($f$) with temperature for the
		Zr$_9$Ni$_{11}$, Zr$_8$Ni$_{21}$ and Zr$_7$Ni$_{10}$ phases are shown in Fig.~\ref{fig:Zr9Ni11_parameter}. A very weak
		temperature dependence of quadrupole frequency for the Zr$_9$Ni$_{11}$ phase is found in this compound. Values of $\omega_Q$ for
		the Zr$_9$Ni$_{11}$ phase shows a linear variation with temperature and it is fitted using the following relation
		\begin{equation}
		\omega_Q(T)=\omega_Q(0)[1-\alpha T]. 
		\label{eqn:T}
		\end{equation}
		The fitted results are found to be $\omega_Q$(0) = 40.5(5) Mrad/s ($V_{zz}$ = 4.5(1)$\times$10$^{21}$ V/m$^2$) and $\alpha$ = 0.7(2)$\times$10$^{-4}$ K$^{-1}$. The quadrupole frequency for the Zr$_8$Ni$_{21}$ phase
		follows $T^{3/2}$ temperature dependence given by the following formula
		\begin{equation}
		\omega_Q(T)=\omega_Q(0)[1-\beta T^{3/2}]. 
		\label{eqn:T32}
		\end{equation}
		The fitting produces values of $\omega_Q(0)$ = 77.1(8) Mrad/s ($V_{zz}$ = 8.6(2)$\times$10$^{21}$ V/m$^2$) and $\beta$ = 1.2(6) $\times$10$^{-6}$ K$^{-3/2}$. In our recent PAC investigation
		in Zr$_8$Ni$_{21}$ \citep{skdeyZr8Ni21}, similar 
		temperature dependence was found.
		
		Temperature evolution of electric quadrupole frequency in metals and intermetallic compounds following a $T$ and $T^{3/2}$ relation are generally 
		found in literature \citep{Christiansen, Schatz}. In most cases, the EFG decreases with temperature. The coefficient $\alpha$ depends on crystal structure and 
		the lattice parameters. A very weak temperature dependence of EFG in Zr$_9$Ni$_{11}$, probably, indicates that lattice parameters or lattice volume does not change
		much with temperature. 
		
		To closely examine the phase transitions, a remeasurement of X-ray powder
		diffraction pattern (Fig.~\ref{fig:Zr9Ni11annealed_XRD}) has been carried out after heating the sample for two days at 1073 K. {Peaks in the XRD spectrum
		were identified (Fig. \ref{fig:Zr9Ni11annealed_XRD}) using ICDD database, 2009.} It is found (Fig.~\ref{fig:Zr9Ni11annealed_XRD}) from XRD analysis using FullProf software \citep{Rodriguez} that the sample after heating gives the phase
		components Zr$_9$Ni$_{11}$ (\citep{StalickZr9Ni11}, PDF Card No.: 00-033-0963), ZrNi (\citep{Bailey}, PDF
		Card No.: 01-072-6477), Zr$_8$Ni$_{21}$ (\citep{Zr8Ni21crystal_structure}, PDF Card No.: 01-071-2622)
		and Zr$_7$Ni$_{10}$ (\citep{Yvon}, PDF Card No.: 01-072-3501) 
		as the constituent phase fractions. The phase
		transformation of Zr$_9$Ni$_{11}$ found at 973 K to ZrNi, therefore, can be supported from XRD measurements also. However, from PAC
		measurement, no phase due to Zr$_9$Ni$_{11}$ has been found at remeasured room temperature (Table \ref{tab:Zr9Ni11_table}). Possibly, in this case, the phase Zr$_9$Ni$_{11}$ is completely
		transformed to ZrNi due to prolonged heating at successive higher temperatures upto 973 K. 
		
		The presence of Zr$_9$Ni$_{11}$, ZrNi, Zr$_8$Ni$_{21}$, ZrNi$_3$ and Zr$_7$Ni$_{10}$ phases in 
		this stoichiometric sample of Zr$_9$Ni$_{11}$ after annealing the sample at 1073 K for 2 days have been observed from TEM/SAED
		measurement (Fig.~\ref{fig:Zr9Ni11annealed_TEM}). The SAED pattern obtained from a region marked by a dotted circle
		in the stoichiometric sample of Zr$_9$Ni$_{11}$ is shown in Fig.~\ref{fig:Zr9Ni11annealed_TEM}. Some of the 
		measured interplanar
		spacings ($d_{hkl}$) from the SAED pattern
		are 3.32 \r{A} and 2.27 \r{A}. These measured $d_{hkl}$ are very close to the (002) and (132) interplanar
		spacings of tetragonal Zr$_9$Ni$_{11}$ (JCPDS $\#$ 33-0963), respectively. This further confirms the presence of Zr$_9$Ni$_{11}$ phase in the sample. 
		Few of the 
		measured $d_{hkl}$ from the SAED pattern
		are  found to be 2.33 \r{A} and 1.70 \r{A} which are very close to the (130) and (150) interplanar
		spacings of orthorhombic ZrNi (JCPDS $\#$ 65-7465), respectively. The presence of ZrNi phase in the sample is thus confirmed. 
		Some of the 
		measured interplanar
		spacings from the SAED pattern, 2.75 \r{A} and 1.70 \r{A}, are found to be very close to the (103) and (-203) interplanar
		spacings of triclinic Zr$_8$Ni$_{21}$ (ICDD PDF Card No.: 01-071-2622), respectively. This confirms the presence of Zr$_8$Ni$_{21}$ phase in the sample. 
		Few of the 
		measured interplanar
		spacings ($d$-spacing) from the SAED pattern
		are 2.32 \r{A} and 1.67 \r{A}. These measured $d_{hkl}$ are very close to the (200) and (112) interplanar
		spacings of hexagonal ZrNi$_3$ (JCPDS $\#$ 65-1968), respectively. Hence, the presence ZrNi$_3$ phase in the sample is confirmed. Measured values of some of the interplanar spacings ($d$-spacing) are 1.77 \r{A}, 2.30 \r{A} and 2.26 \r{A} which are found to be very close to the (151), (004) and (023) interplanar spacings of Zr$_7$Ni$_{10}$ (ICDD PDF Card No.:01-072-3501), respectively. Therefore, the presence of Zr$_7$Ni$_{10}$ phase in this sample is confirmed.

		\begin{table}{h}
				\begin{center}
			\caption{\label{tab:Hf9Ni11_table}Results of PAC measurements in Hf$_9$Ni$_{11}$}
			\scalebox{0.85}{
				\begin{tabular}{ccccccc}
					\hline
					Temperature (K)  &Component    & $\omega_Q$ (Mrad/s)     & $\eta$     & $\delta$($\%$)   & $f$($\%$)      & Assignment   \\ 
					\hline  	
					298             &1        & 25.7(3)                 & 0.37(3)          & 12(1)            & 81(3)      &   HfNi      \\   
					&2      & 39.5(5)                 & 0.64(4)          & 0                & 19(3)         &    Hf$_9$Ni$_{11}$ \\   \\
					
					373               &1     & 22(1)                 & 0.39(7)          & 0              & 55(3)    &  HfNi         \\   
					&2      & 40.9(9)                 & 0.59(7)          & 0                & 45(3)  &        Hf$_9$Ni$_{11}$  \\    \\
					
					473              &1      & 19.3(9)                 & 0.26(8)          & 0                & 64(3)    &   HfNi          \\ 
					&2      & 40.0(8)                 & 0.59(6)          & 0                & 36(3)  &        Hf$_9$Ni$_{11}$  \\    \\
					
					573              &1      & 18(1)                 & 0.25(fixed)          & 0                & 76(3)   &    HfNi        \\   
					&2      & 41(2)                 & 0.59(12)          & 0                & 24(3)  &        Hf$_9$Ni$_{11}$  \\    \\
					
					673              &1      & 18.7(5)              & 0.18(7)          & 0              & 67(3)  &      HfNi       \\  
					&2      & 41.7(6)                 & 0.58(3)          & 0                & 33(3)  &      Hf$_9$Ni$_{11}$  \\ \\
					
					773              &1      & 18.4(5)                 & 0.30(4)        & 0               & 61(3)  &      HfNi    \\  
					&2      & 42.6(4)                & 0.57(2)                & 0                & 39(3)   &       Hf$_9$Ni$_{11}$     \\  \\
					
					873              &1      & 17.2(4)                 & 0.19(5)          & 0                & 68(3)    &     HfNi   \\   
					&2      & 41.9(4)                & 0.54(2)        & 0                & 32(3)  &             Hf$_9$Ni$_{11}$\\ \\        
					
					298$^a$        &1    & 24.3(4)                 & 0.38(2)         & 3(1)                & 75(3)         &  HfNi\\ 
					&2   & 42.5(5)                & 0.68(3)          & 0                & 25(3)    &  Hf$_9$Ni$_{11}$  \\     
					\hline                          
				\end{tabular}}
					\end{center}
			\begin{flushleft}
			{$^a$ after measurement at 873 K}\\
			\end{flushleft}
		\end{table}

			\begin{figure*}[t!]
				\centering
				\begin{subfigure}[t]{0.40\textwidth}
					\centering
					\includegraphics[scale=.38]{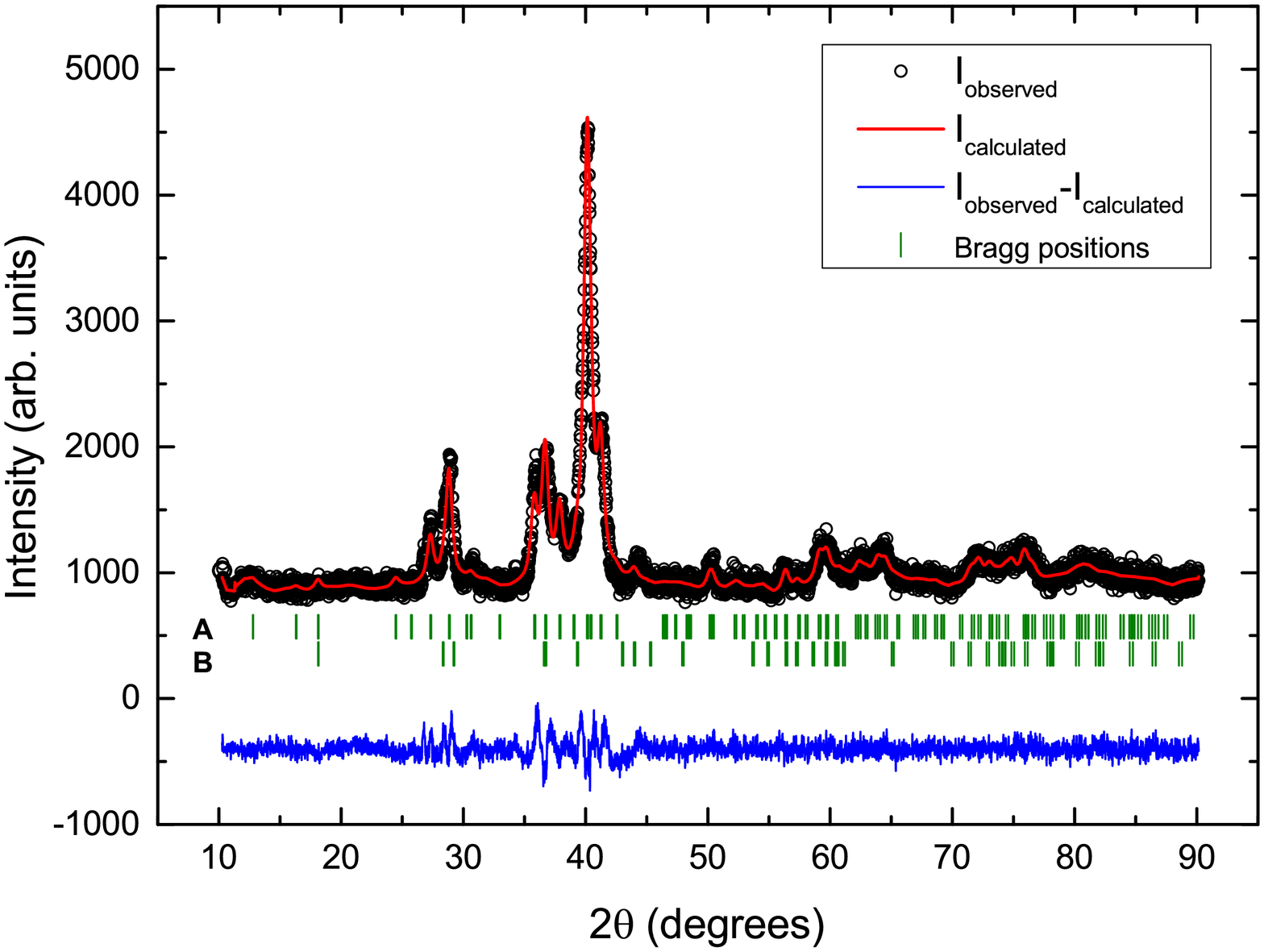}
				\end{subfigure}
				\hspace{2.7cm}
				\begin{subfigure}[t]{0.4\textwidth}
					\centering
					\includegraphics[scale=.18]{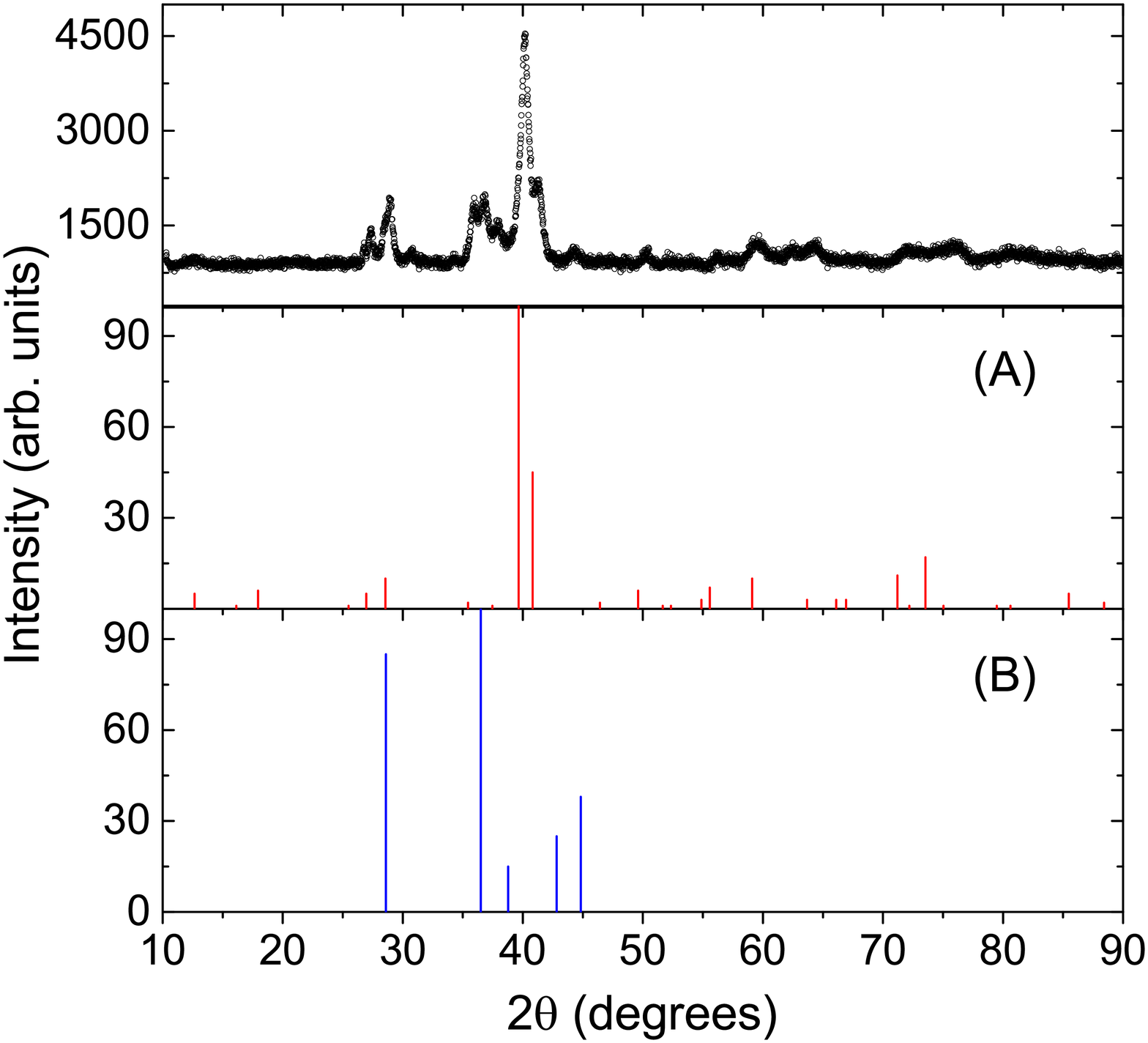}
				\end{subfigure}
				\caption{Figure in the left shows the background subtracted XRD powder in the stoichiometric sample of Hf$_9$Ni$_{11}$. The line represents the fit to the measured data. 
					The vertical bars A, B denote the Bragg angles corresponding to Hf$_9$Ni$_{11}$ and HfNi, respectively. 
					The bottom line shows the difference between the observed and the fitted pattern. {Figure in the right shows a comparison of ICDD pattern of Zr$_9$Ni$_{11}$ (A), HfNi (B) with the experimental XRD pattern of the as-cast Hf$_9$Ni$_{11}$ sample.  In the database, no ICDD pattern file was found for Hf$_9$Ni$_{11}$. Therefore, ICDD pattern of isostructural Zr$_9$Ni$_{11}$ \citep{Liu,KrikpatrickLarson,PAnash} has been used for identifying Hf$_9$Ni$_{11}$ phase.}}
				\label{fig:Hf9Ni11ascast_XRD}
			\end{figure*}
		
%
%
%
%
		
		\begin{figure}[t!]
			\begin{center}
				\includegraphics[width=0.6\textwidth]{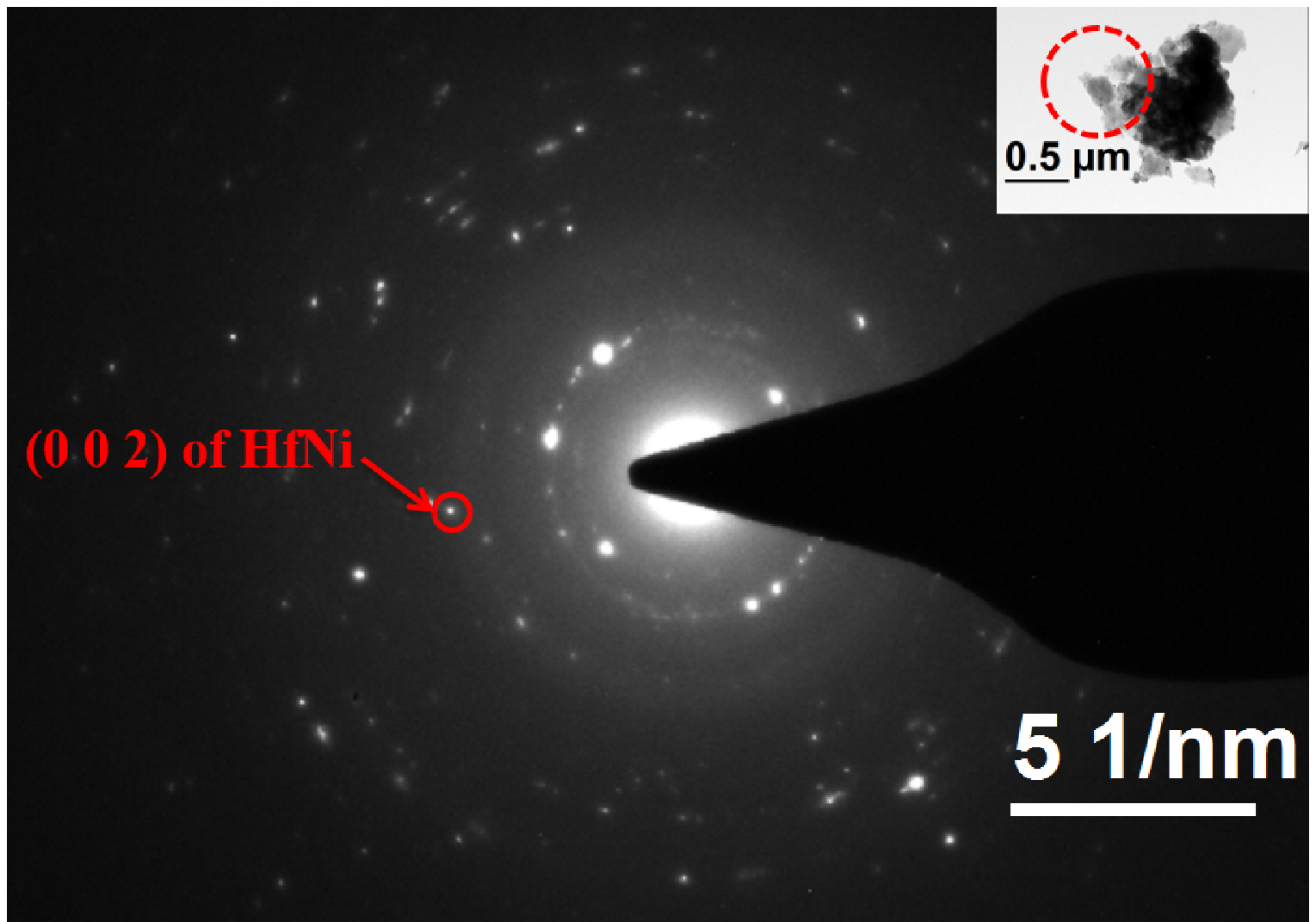}
			\end{center}
			\caption{			\label{fig:Hf9Ni11ascast_TEM}Selected area electron diffraction pattern from Hf$_9$Ni$_{11}$ particle shown in the inset.}
		\end{figure}

		\begin{figure*}[t!]
			\begin{center}
				\includegraphics[width=0.8\textwidth]{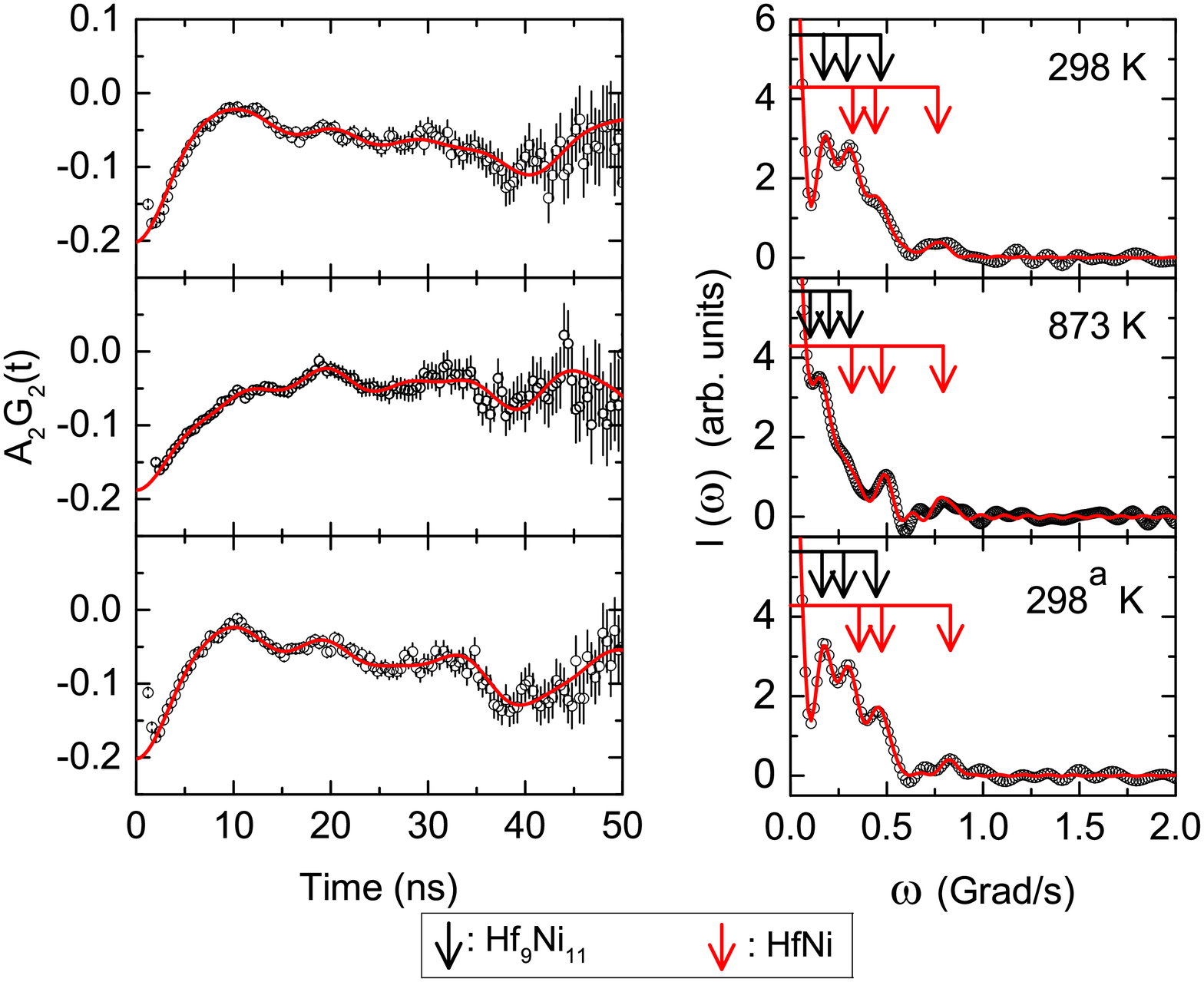}
			\end{center}
			\caption{			\label{fig:Hf9Ni11_spectra}PAC spectra in the stoichiometric sample of Hf$_9$Ni$_{11}$ at different
					temperature. Left panel
					shows the time spectra and the right panel shows
					the corresponding Fourier cosine transforms. The PAC spectrum at room temperature designated by 298$^a$ K is taken after the
					measurement at 873 K. }
		\end{figure*} 
		
		\begin{figure}[t!]
			\begin{center}
				\includegraphics[width=0.45\textwidth]{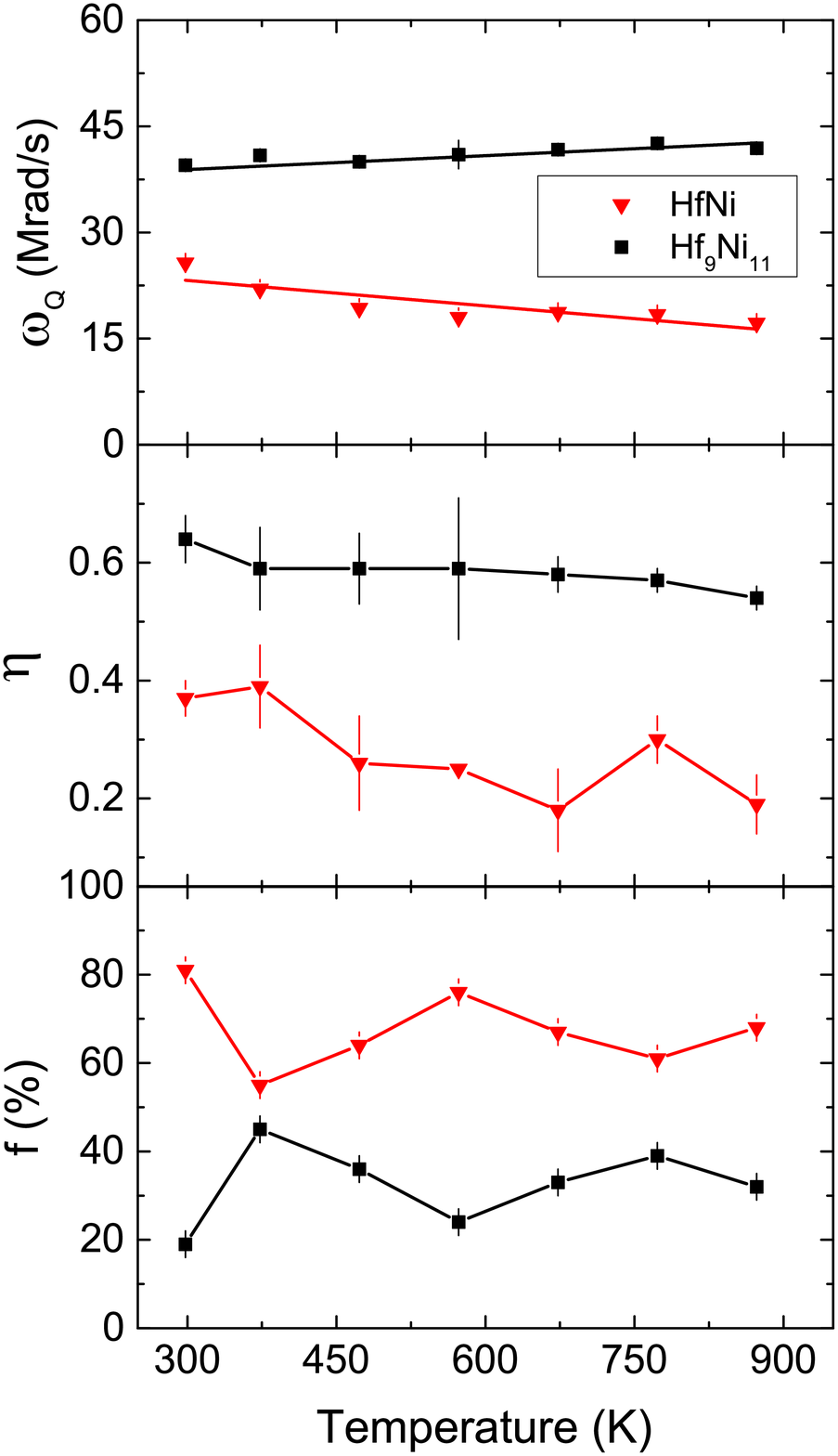}
			\end{center}
			\caption{			\label{fig:Hf9Ni11_parameter}Variations of quadrupole frequency ($\omega_Q$), asymmetry parameter ($\eta$) and site fraction $f$(\%)
					with temperature for the components of Hf$_9$Ni$_{11}$ and HfNi.}

		\end{figure}
		
		\subsection{Hf$_9$Ni$_{11}$}
		The powder X-ray diffraction pattern in Hf$_9$Ni$_{11}$ is shown in Fig.~\ref{fig:Hf9Ni11ascast_XRD}. {Presence
		of Hf$_9$Ni$_{11}$ and HfNi phases have been found (Fig. \ref{fig:Hf9Ni11ascast_XRD}) in the XRD spectrum by comparing with the ICDD PDF Card No.: 00-033-0963 of Zr$_9$Ni$_{11}$ and JCPDS $\#$ 47-1414 of HfNi, respectively.} Analysis of
		the X-ray powder pattern (Fig.~\ref{fig:Hf9Ni11ascast_XRD}) was
		carried out by FullProf software package \citep{Rodriguez} using the known crystallographic data 
		of body-centered tetragonal Hf$_9$Ni$_{11}$ \citep{KrikpatrickLarson,Liu,PAnash} and base-centered orthorhombic HfNi \citep{KrikpatrickLarson,Bailey,PAnash} phases. 
		The presence of HfNi phase in
		this stoichiometric sample of Hf$_9$Ni$_{11}$ has been observed from TEM/SAED measurement (Fig.~\ref{fig:Hf9Ni11ascast_TEM}). Selected area electron
		diffraction (SAED) pattern obtained from a region marked by a dotted circle in the stoichiometric sample of Hf$_9$Ni$_{11}$ is shown in Fig.~\ref{fig:Hf9Ni11ascast_TEM}. One of the 
		measured interplanar
		spacing from the SAED pattern
		is 2.03 \r{A} which is found to be very close to the (002) interplanar
		spacing of orthorhombic HfNi (JCPDS $\#$ 47-1414). This further confirms the presence of HfNi phase in the sample. The phase Hf$_9$Ni$_{11}$ 
		could not be identified from SAED pattern in the stoichiometric sample of Hf$_9$Ni$_{11}$ due to non-availability of X-ray diffraction data of
		interplanar spacings ($d_{hkl}$) and corresponding crystallographic planes ($hkl$) for Hf$_9$Ni$_{11}$.  
		
		The PAC spectrum at room temperature in Hf$_9$Ni$_{11}$ is shown in Fig.~\ref{fig:Hf9Ni11_spectra}.
		An analysis of the spectrum produces one predominant 
		component ($\sim$81\%) with values of $\omega_Q$ = 25.7(3) Mrad/s, $\eta$ = 0.37(3)
		and $\delta$ = 12(1)\% and a minor component ($\sim$19\%) with
		values of $\omega_Q$ = 39.5(5) Mrad/s, $\eta$ = 0.64(4), $\delta$ = 0. Comparing
		these values with the PAC results in Zr$_9$Ni$_{11}$ at room temperature (Table \ref{tab:Zr9Ni11_table}), the
		minor component can be assigned to Hf$_9$Ni$_{11}$. Since the crystal structures of Zr$_9$Ni$_{11}$ and Hf$_9$Ni$_{11}$
		are the same, similar values of $\omega_Q$ and $\eta$ in the two compounds
		are expected. The predominant component, however, can be
		attributed to HfNi by comparing these results with the
		previous reported results in HfNi \citep{Yaar}. In the Hf-Ni phase diagram \citep{PAnash}, nine intermediate phases have been found where 
		the phases Hf$_2$Ni$_7$ and HfNi melt congruently, and
		seven other phases of Hf-Ni binary system, namely, Hf$_9$Ni$_{11}$, $\alpha$/$\beta$-HfNi$_3$, Hf$_8$Ni$_{21}$, Hf$_3$Ni$_7$, Hf$_7$Ni$_{10}$, Hf$_2$Ni,
		HfNi$_5$ are formed peritectically \citep{PAnash}. The Hf$_9$Ni$_{11}$ phase is formed by peritectic reaction
		from the liquid melt and HfNi at 1613 K \citep{Svechnikov,Kejun}. Kirkpatrick and Larson \citep{KrikpatrickLarson} reported that the Hf-Ni system
		is analogous to Zr-Ni system. Unlike Zr$_9$Ni$_{11}$,
		a predominant component due to HfNi is observed in Hf$_9$Ni$_{11}$. This is,
		probably, due to the fact that in Hf$_9$Ni$_{11}$ PAC sample, a significant mass loss ($\sim$5.5\%) of the sample was observed during its preparation
		in argon arc furnace. This mass loss can be considered as
		due to loss of Ni which has a much lower melting and
		boiling point compared to the other constituent Hf. So,
		the stoichiometry of the arc melted sample was changed and
		a Ni deficient compound was produced by melting. Since,
		the stoichimetry of Hf$_9$Ni$_{11}$ (Hf/Ni ratio 1:1.22) and HfNi
		are not very different, the production of HfNi phase in
		the sample is expected. In this case, the mass loss
		of the sample was such that its stoichiometry comes
		closer to HfNi. 
		
		The PAC measurements at higher temperatures have also been
		performed in Hf$_9$Ni$_{11}$. It is found that at 373 K,
		the fraction due to Hf$_9$Ni$_{11}$ enhances to 45\% at the expense
		of HfNi (Table \ref{tab:Hf9Ni11_table}). In the temperature range 373-873 K,
		the component fractions do not change much (Fig.~\ref{fig:Hf9Ni11_parameter}). 
		
		After heating the sample gradually up to 873 K,
		a re-measurement at room temperature was carried out.
		The results are found to be almost the same as found before
		heating. The frequency distribution width for the
		predominant component, however, decreases (Table \ref{tab:Hf9Ni11_table})
		as expected due to annealing of the sample. In this
		case, the values of quadrupole frequencies for the
		two components are found to change slightly. In the two cases,
		there are large changes in frequency distribution widths which
		probably affect the results. But unlike Zr$_9$Ni$_{11}$, no change
		in phases due to heating is observed in Hf$_9$Ni$_{11}$ and the phases
		are found to be stable up to 873 K.
		
		The variations of quadrupole frequency and asymmetry
		parameters of two phases found in the stoichiometric sample of Hf$_9$Ni$_{11}$ with temperature are shown in Fig.~\ref{fig:Hf9Ni11_parameter}.
		It is found that the temperature dependence of quadrupole
		frequency for the Hf$_9$Ni$_{11}$ phase is very weak which is similar to the temperature
		dependence of quadrupole frequency in Zr$_9$Ni$_{11}$. But, unlike Zr$_9$Ni$_{11}$, 
		$\omega_Q$ for Hf$_9$Ni$_{11}$ is found to increase with temperature. Although, the $\omega_Q$ for this phase is found to
		evolve with temperature following a linear temperature dependent relation (Eqn. \ref{eqn:T}). A fitting to the Eqn. \ref{eqn:T} produces
		values of $\omega_Q$(0) = 36.9(9) Mrad/s ($V_{zz}$=4.1(1)$\times$10$^{21}$ V/m$^2$) and $\alpha$ = -1.9(3)$\times$10$^{-4}$ K$^{-1}$. Similar unusual temperature 
		dependence of quadrupole frequency was also found in the literature \citep{Wodniecki}. Here also, a weak temperature dependence of EFG indicates that crystal 
		parameters do not change much with temperature. The $\omega_Q$ for the predominant HfNi component is also found to vary linearly with temperature
		following Eqn. \ref{eqn:T}. The fitted results are found to be $\omega_Q$(0) = 27(2) Mrad/s ($V_{zz}$ = 3.0(2)$\times$10$^{21}$ V/m$^2$) and
		$\alpha$=4(1)$\times$10$^{-4}$ K$^{3/2}$.

		\begin{table}
			\caption{\label{tab:Zr9Ni11_Hf9Ni11_structure}The parameters of the Zr$_9$Ni$_{11}$ and Hf$_9$Ni$_{11}$ structure.
				The lattice constants are given in $\AA$.}
			\scalebox{0.88}{
					\begin{tabular}{cccc}
						\hline
	&Our calculated results   & Previous experimental & Present experimental \\ &(WIEN 2k) \citep{PBlaha} & results (XRD) \citep{Glimois,KrikpatrickLarson,Akihiro,Kosorukova,Liu} 
						& results (XRD) \\ [1.5ex]
						\hline  
Zr$_9$Ni$_{11}$ & & \\	
						$a$ &9.88 &9.88(1) \citep{Glimois}, 9.9778 \citep{Kosorukova} &9.881(2)\\
						$c$ &6.48 &6.61(1) \citep{Glimois}, 6.5809 \citep{Kosorukova} &6.613(1)\\
						$B$ [GPa] &144 & \\  \\
Hf$_9$Ni$_{11}$ & & \\
						$a$ &9.89 &9.79\citep{KrikpatrickLarson} &9.801(2)\\
						$c$ &6.44 &6.53\citep{KrikpatrickLarson} &6.532(2)\\
						$B$ [GPa] &161 &\\ 
						\hline                         
					\end{tabular}}
			\end{table}

		\begin{table}
			\caption{\label{tab:DFT_Zr9Ni11_Hf9Ni11}The calculated and experimental values of EFG 
				(in units of 10$^{21}$ V/m$^2$) and asymmetry parameter ($\eta$) for Zr$_9$Ni$_{11}$ and Hf$_9$Ni$_{11}$.}
								\scalebox{0.8}{
				\begin{tabular}{cccccc}	
					\hline	
Compound  &Lattice site    & Calculated EFG  & Calculated $\eta$ & Measured EFG  & Measured $\eta$  \\ [1.5ex]
					\hline  
Zr$_9$Ni$_{11}$ ($^{181}$Ta probe)   &Zr1 2$a$ &25.9 &0 & &\\
					&Zr2 8$h$   &-8.4 &0.30 & &\\
					&Zr3 8$h$  &-5.6 &0.92 &4.6(1) & 0.80(10) \\ \\
Hf$_9$Ni$_{11}$ ($^{181}$Ta probe)   &Hf1 2$a$  &20.7 &0 & &\\
					&Hf2 8$h$ &-8.1 &0.39 & &\\
					&Hf3 8$h$   &5.0 &0.72 &4.7(1) &0.68(3)\\
					\hline                         
				\end{tabular}}
		\end{table}	

				\begin{figure*}[t!]
					\begin{center}
						\includegraphics[width=0.9\textwidth]{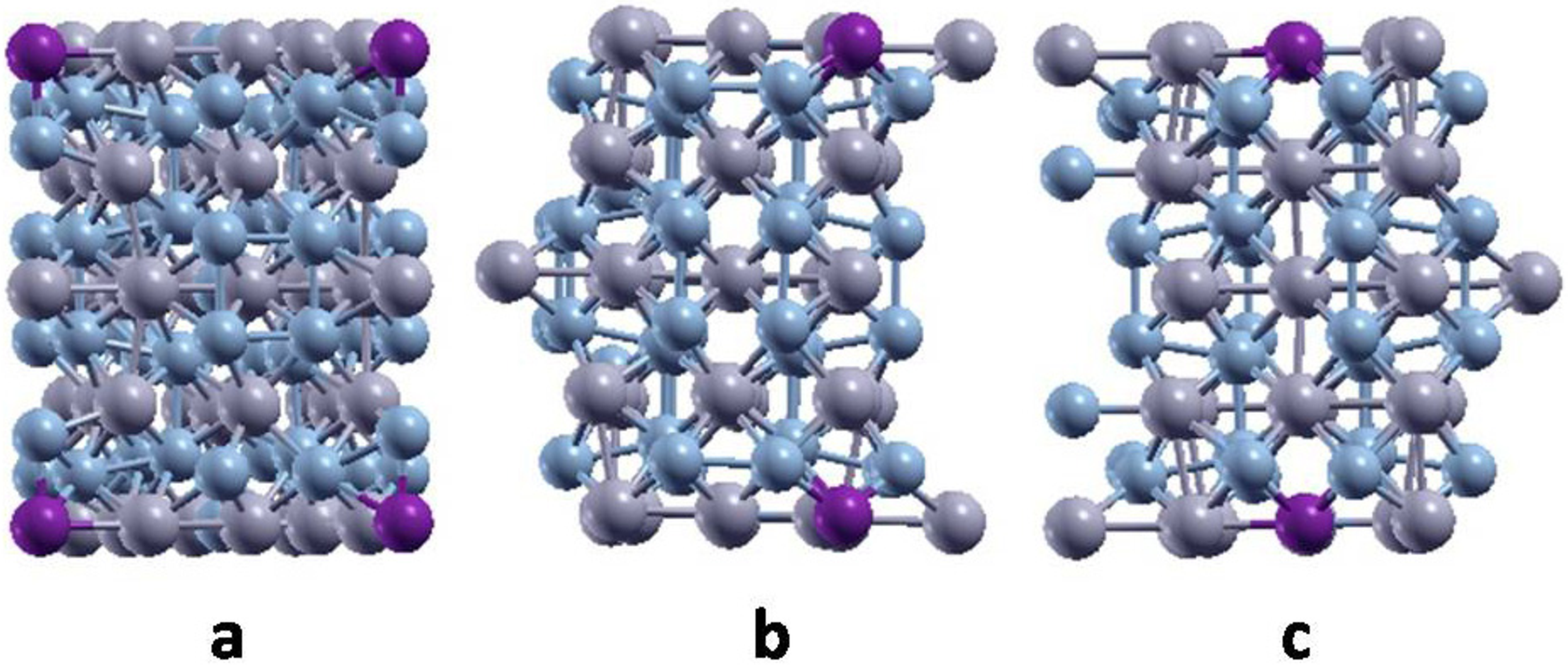}
					\end{center}
					\caption{\small{Models of cells used in DFT study. The Hf, Ni and Ta atoms are denoted by grey, blue and purple spheres, respectively.}}
					\label{Hf9Ni11_cells}
				\end{figure*}
				\section{Ab initio calculations}
				The ab initio density functional theory (DFT) calculations were performed to compare with the
				experimental results by all-electron full potential (linearized)
				augmented plane waves plus local orbitals [FP-(L)APW+lo] method, as implemented in WIEN2k \citep{PBlaha}. The exchange-correlation potential was calculated using the generalized gradient approximation (GGA) with the parameterization of Perdew-Burke-Ernzerhof (PBE) \citep{Perdew}.
				The cut-off parameter $R_{mt}K_{max}$ for limiting the number of plane waves, where $R_{mt}$ is the
				smallest value of all atomic sphere radii and $K_{max}$ is the largest reciprocal lattice vector used in the
				plane wave expansion, was set to 7.0. The Brillouin zone integration was achieved via a tetrahedron method \citep{Blochl}. Taking
				into consideration both the accuracy and the efficiency of the calculations, we have selected a 10 $\times$ 10 $\times$ 10 point
				mesh to sample the entire Brillouin-zone (BZ), yielding 143 points in the irreducible Brillouin-zone. The structure
				was relaxed according to Hellmann-Feynman forces calculated at the end of each self-consistent cycle, until the
				forces acting on all atoms were less than 0.03 eV/$\AA$ (2 mRy/a.u.). During this relaxation, the cell volume was kept
				fixed to its experimental value. Then the theoretical equilibrium volume was determined by fixing the atomic positions
				to their optimized values and further keeping $c/a$ ratio fixed. A series of calculations was carried out, changing
				the volume within $\pm$ 5\% of its experimental value and calculating the total energy as its function. Finally,
				the $c/a$ ratio was optimized by changing it within $\pm$ 2\% of its experimental value while keeping the optimized
				volume fixed. The convergence criterion for achieving self-consistency was that the integrated
				charge difference between last two iterations to be smaller than 5 $\times$ 10$^{-5}$ electron. 
				
				We used the optimal structural parameters to construct 1 $\times$ 1 $\times$ 2 supercell and then replaced each of the Zr (Hf) nonequivalent host sites by
				a Ta atom successively (Figure \ref{Hf9Ni11_cells}, \citep{Kokalj}). The point group symmetry around the Ta atom remained the same as around the original atom,
				but the number of non-equivalent positions enlarged to 18-50, which increased the complexity of the calculations. We
				used 36-64 $k$ points in the irreducible wedge of the Brillouin zone for these calculations.
				
				Both Zr$_{9}$Ni$_{11}$ and Hf$_9$Ni$_{11}$ have tetragonal $I4/m$ type crystal structure (space group
				number 87). For Zr$_9$Ni$_{11}$, $P4/m$ structure (space group no 83)
				was reported by Stalick et al. \citep{StalickZr9Ni11} only, but other authors found $I4/m$ structure \citep{Glimois, Akihiro, Kosorukova} including two recent results
				by Matsuyama et al. \citep{Akihiro} and Kosorukova et al. \citep{Kosorukova}. We have found $I4/m$ crystal structure from XRD (Table \ref{tab:Zr9Ni11_Hf9Ni11_structure},
				Fig.~\ref{fig:Zr9Ni11ascast_XRD}, \ref{fig:Zr9Ni11annealed_XRD}, \ref{fig:Hf9Ni11ascast_XRD}) and TEM/SAED (Figure \ref{fig:Zr9Ni11ascast_TEM}, \ref{fig:Zr9Ni11annealed_TEM}) studies
				in our samples. So, for DFT calculations, we have considered $I4/m$ crystal structure for both Zr$_9$Ni$_{11}$ and Hf$_9$Ni$_{11}$. 
				This structure contains 20 atoms in the primitive cell and three nonequivalent Hf (Zr) crystallographic
				positions. The theoretically optimized structure parameters together with the experimental values obtained
				from earlier \citep{Glimois,Kosorukova} and present X-ray diffraction measurements are shown in Table \ref{tab:Zr9Ni11_Hf9Ni11_structure}. We can
				see that $a$ lattice constants
				overestimate the experimental ones for about 1\%, while the $c$ lattice constants underestimate
				the measured ones for about 2\%. The bulk moduli $B$, obtained by
				fitting the data to the Murnaghan's equation of state \citep{Murnaghan} are also given in Table \ref{tab:Zr9Ni11_Hf9Ni11_structure}.
				
				The electric field gradient (EFG) tensor $V_{ij}$ was calculated from the obtained charge density using the method developed in reference \citep{Herzig}. The usual
				convention is to designate the largest component of the EFG tensor as $V_{zz}$. The asymmetry parameter $\eta$ is then
				given by $\eta=(V_{xx}-V_{yy})/V_{zz}$, where $|V_{zz}|\ge |V_{yy}|\ge |V_{xx}|$. The calculated and experimental EFGs at the Ta probe positions in the 
				Hf$_9$Ni$_{11}$ and Zr$_9$Ni$_{11}$ compounds, along with the corresponding asymmetry parameter values, are given in Table \ref{tab:DFT_Zr9Ni11_Hf9Ni11}. 
				It can be observed that the EFG values for Ta at 2$a$ position are not very different in Hf$_9$Ni$_{11}$
				and Zr$_9$Ni$_{11}$. This is the position with largest EFG. The EFGs for Ta at the other two nonequivalent
				Hf(Zr) positions, 8$h$, are also similar in both compounds. We see that the calculated result for EFG and $\eta$ at the Ta probe replacing Zr3
				atom (-5.6 $\times$ 10$^{21}$ V/m$^2$ and 0.92) is in agreement with the corresponding values measured at 77 K (4.6 $\times$ 10$^{21}$ V/m$^2$ and 0.80), thus confirming that the mentioned component of the
				measured PAC spectra originates from Zr$_9$Ni$_{11}$. The sign of $V_{zz}$ can not be determined by PAC measurements. Therefore, absolute values of $V_{zz}$
				have been compared. In the case of Hf$_9$Ni$_{11}$, the calculated results of EFG (5.0 $\times$ 10$^{21}$ V/m$^2$) and $\eta$ (0.72) are in better agreement with the
				measured values of EFG (4.4 $\times$ 10$^{21}$ V/m$^2$) and  $\eta$ (0.64) at room temperature.
				It is interesting to notice that remeasured values at room temperature after heating
				(4.7 $\times$ 10$^{21}$ V/m$^2$ and 0.68) exhibit even better agreement with the calculated ones. 
				Possible explanation is that the heating resulted in the better
				structure arrangement, which is now closer to the ideal one.
				
				\section{Conclusion}
				We have performed PAC measurements and DFT calculations to address the phase components, and phase stability in Zr$_9$Ni$_{11}$ and Hf$_9$Ni$_{11}$. In both these compounds, multiphase components have been found which are assigned by comparing with the results from DFT calculations. Three non-equivalent Zr/Hf sites are found from calculations while only one site occupation (Zr3/Hf3) by the probe is obtained experimentally.
				In Zr$_9$Ni$_{11}$ stoichiometric sample, up to 673 K, the phase due to the Zr$_9$Ni$_{11}$ is found to be predominant, with
				the minor phase belonging to Zr$_8$Ni$_{21}$. Above 673 K,
				the Zr$_9$Ni$_{11}$ phase becomes unstable and Zr$_7$Ni$_{10}$ appears. 
				At 973 K, the phase due to ZrNi is produced as a dominating phase. This phase has also been found to be predominant when remeasured at room temperature after 
				measurement at 973 K. 
				
				In the stoichiometric Hf$_9$Ni$_{11}$ sample, the phase due to Hf$_9$Ni$_{11}$
				is produced as a minority phase and the HfNi phase is found to
				be predominantly produced. From our PAC measurements, isostructurality of Zr$_9$Ni$_{11}$ and Hf$_9$Ni$_{11}$ has been found. The phases produced in Hf$_9$Ni$_{11}$ are found to be stable up to 873 K.
				
				In both Zr$_9$Ni$_{11}$ and Hf$_9$Ni$_{11}$, very weak temperature dependence
				of electric field gradients (EFG) have been observed which indicates that the lattice
				parameters do not change much with temperature. The calculated results of EFG and asymmetry parameter ($\eta$) at the
				Ta impurity atom positions, Zr3 8$h$ in Zr$_9$Ni$_{11}$ and Hf3 8$h$ in Hf$_9$Ni$_{11}$, are found
				to be in good agreement with our experimental results, thus confirming the origin of the components of our interest in the PAC spectra.  
				
				\vspace{1 cm}
				{\hspace{ -0.4 cm}}{\bf Acknowledgement}
				\vspace{0.5cm}
				
The authors acknowledge with thanks A. Karmahapatra and S. Pakhira of Saha Institute of Nuclear Physics, Kolkata for their helps in XRD
measurements and data analysis. The present work is supported by the Department of Atomic Energy, Government of India through the Grant no. 12-R\&D-SIN-5.02-0102. J. Belo$\check{\text{s}}$evi\'{c}-$\check{\text{C}}$avor acknowledges support by The Ministry of Education, Science and Technological Development of the Republic of Serbia through the project no. 171001.
				
\bibliographystyle{}

\begin{thebibliography}{56}
					
\bibitem{Pang} S.J. Pang, T. Zhang, K. Asami, Mat. Trans. {\bf 43}, 1771 (2002). 
					
\bibitem{Ivanchenko} V. Ivanchenko, T. Kosorukova, M. Samohin, S. Samohin, Yu. Butenko, Patent of Ukraine on useful model 26254 (2007). 
					
\bibitem{Haour} G. Haour, F. Mollard, B. Lux, and I. G. Wright, Z. Metallk. {\bf 69}, 149 (1978).
					
\bibitem{Koch} P. Nash, in High-Temperature Ordered Intermetallic Alloys, edited by C. C. Koch, C. T. Liu, and N. S. Stoloff (Mater. Res. Soc. Symp. Proc. 39, Pittsburgh, PA, 1985), p. 423.
					
\bibitem{Ruiz} F.C. Ruiz, E.B. Castro, S.G. Real, H.A. Peretti, A. Visintin, W.E. Triaca, Int. J. Hydrogen Energy {\bf 33}, 3576 (2008).
  
\bibitem{FCRuiz}F. C. Ruiz, E. B. Castro, H. A. Peretti, A. Visitin, Int. J. Hydrogen Energy {\bf 35}, 9879 (2010).
					
\bibitem{Nei} J. Nei, K. Young, R. Regmi, G. Lawes, S.O. Salley, K.Y.S. Ng, Int. J. Hydrogen Energy {\bf 37}, 16042 (2012).
					
\bibitem{Joubert} J.M. Joubert, M. Latroche, A. Percheron-Gu{\'e}gan, J. Alloys Compd. {\bf 231}, 494 (1995).
					
\bibitem{Zhang}X.-Y. Song, X.-B. Zhang, Y.-Q. Lei, Z. Zhang, Q.-D. Wang, Int. J. Hydrogen Eng. {\bf 24}, 455 (1999).
					
\bibitem{Young} Kwo Young, Taihei Ouchi, Michael A. Fetcenko, Willy Mays, Benjamin Reichman, Int. J. Hydrogen Energy {\bf 34}, 8695 (2009).
					
\bibitem{Ouchi} Young K, Ouchi T, Liu Y, Reichman B, Mays W, Fetcenko MA., J Alloys Compd. {\bf 480}, 521 (2009).
					
\bibitem{Fetcenko}K. Young, J. Nei, T. Ouchi, M.A. Fetcenko, Journal of Alloys and Compounds {\bf 509}, 2277 (2011).
					
\bibitem{Kwo} Kwo-hsiung Young and Jean Nei, Materials {\bf 6}, 4574 (2013).
					
\bibitem{Akihiro} Akihiro Matsuyama, Hironori Mizutani, Takumi Kozuka, Hiroshi Inoue, J. Alloys Compd. {\bf 714}, 467 (2017)
					
\bibitem{ProvenzanoZr9Ni11} V. Provenzano, R.D. Shull, R.M. Waterstrat, L.H. Bennett, E. Della Torre and H. Seyoum, IEEE Trans. Magnetics {\bf 46}, 502 (2010).
					
\bibitem{Glimois}J.L. Glimois, C. Becle, G. Develey, J.M. Moreau, J. Less-Common Met. {\bf 64}, 87 (1979).
					
\bibitem{KrikpatrickLarson}M.E. Kirkpatrick, W.L. Larsen, Trans. Am. Soc. Met. {\bf 54}, 580 (1961).
					
\bibitem{Shadangi}S.K. Shadangi, S.C. Panda and S. Bhan, Acta Crystallogr. B {\bf 38}, 2092 (1982).
					
\bibitem{Xueyan} Xueyan Song, Ze Zhang, Xiaobin Zhang, Yongquan Lei, and Qidong Wang, J. Mater. Res. {\bf 14}, 1279 (1999).
					
\bibitem{Nash} P. Nash, C.S. Jayanth, Bulletin of Alloy Phase Diagrams {\bf 5}, 144 (1984). 
					
\bibitem{Kosorukova} T. Kosorukova, V. Ivanchenko, G. Firstov and H. No\"{e}l, Solid State Phenomena {\bf 194}, 14 (2013).
					
\bibitem{Bhan} S.C. Panda, S. Bhan, J. Less-Common Met. {\bf 34}, 344 (1974).
					
\bibitem{Panda}S.C. Panda and S. Bhan, Z. Metallk. {\bf 64}, 793 (1973) H11.
					
\bibitem{Liu} J.L. Liu, L.L. Zhu, X.M. Huang, G.M. Cai and Z.P. Jin, CALPHAD: Computer Coupling of Phase Diagrams and Thermochemistry {\bf 58}, 160 (2017).
					
\bibitem{PAnash}  P. Nash, A. Nash, Bulletin of Alloy Phase Diagrams {\bf 4}, 250 (1983).
					
\bibitem{Ghosh} G. Ghosh, J. Mat. Res. {\bf 90}, 598 (1994).
					
\bibitem{Bsenko} Lars Bsenko, J. Less-Common Metals {\bf 63}, 171 (1979).

\bibitem{Xiaoma}Xiaoma Tao, Pei Yao, Wenwang Wei, Hongmei Chen, Yifang Ouyang, Yong Du, Yuan Yuan, Qing Peng, Journal of Alloys and Compounds {\bf 752}, 412 (2018).
					
\bibitem{StalickZr9Ni11} J.K. Stalick, L.A. Bendersky and R.M. Waterstrat, J. Phys. Cond. Matt. {\bf 20}, 285209 (2008).
					
\bibitem{skdeyZr8Ni21} S.K. Dey, C.C. Dey, S. Saha, J. Belo$\check{\text{s}}$evi\'{c}-$\check{\text{C}}$avor, Intermetallics {\bf 84}, 112 (2017).
					
\bibitem{CCDeyPhysica} C. C. Dey and S. K. Srivastava, Physica B {\bf 427}, 126 (2013).
					
\bibitem{Marszalek} M. Marszalek, H. Saitovitch, P.R.J. Silva, Z. Naturforsch. {\bf 55a}, 49 (2000).
					
\bibitem{SKDey_JACOM} S.K. Dey, C.C. Dey, S. Saha, J. Belo$\check{\text{s}}$evi\'{c}-$\check{\text{C}}$avor, D. Toprek, J. Alloys Compd. {\bf 723}, 425 (2017).
					
\bibitem{Firestone} R.B. Firestone, V.S. Shirley (Eds.), Table of Isotopes, 8th ed., John Wiley and Sons, New York, 1996.
					
\bibitem{Schatz} G. Schatz, A. Weidinger, Nuclear condensed matter physics, Nuclear Methods and Application, John Wiley and Sons, Chichester, New York, Brisbane, Toronto, Singapore, 1996, p. 63 (chapter 5).
					
\bibitem{Zacate}M. Zacate and H. Jaeger, Defect Diffus. Forum {\bf 311}, 3 (2011).  

\bibitem{Butz}T. Butz, A. Lerf, Phys. Lett. A {\bf 97}, 217 (1983).
					
\bibitem{pramana} C.C. Dey, Pramana {\bf 70}, 835 (2008).
					
\bibitem{Rodriguez}  J. Rodr{\'i}guez-Carvajal, Physica B {\bf 192}, 55 (1993).
					
\bibitem {Yvon}  J.-M. Joubert, R. {\v C}ern{\'y}, K. Yvon, M. Latroche, A.P. Gu{\'e}gan,  Acta Crystallogr. C {\bf 53}, 1536 (1997).
					
\bibitem{skdeyZr7Ni10} S.K. Dey, C.C. Dey, S. Saha, G. Bhattacharjee, D. Banerjee, J. Belo$\check{\text{s}}$evi\'{c}-$\check{\text{C}}$avor, D. Toprek (arXiv id: 1809.08114).
					
\bibitem{CCDeyJPCS} C. C. Dey, Rakesh Das and S. K. Srivastava, J. Phys. Chem. Solids {\bf 82}, 10 (2015).
					
\bibitem{Christiansen} W. Witthuhn and E. Engel, in Hyperfine Interactions of Radioactive Nuclei, Ed. by J. Christiansen (Springer-Verlag Berlin Heidelberg New York Tokyo, 1983) P. 205.
					
\bibitem{Bailey}M.E. Kirkpatrick, D.M. Bailey and J.F. Smith, Acta Cryst. {\bf 15}, 252 (1962).
					
\bibitem{Zr8Ni21crystal_structure} J.M. Joubert, R. cern{\'y}, K. Yvon, Z. Kristallogr-New Cryst. Struct. {\bf 213} 227 (1998).
					
\bibitem{Yaar} I. Yaar, D. Cohen, I. Halevy, S. Kahane, H. Ettedgui, R. Aslanov, Z. Berant, Hyperfine Interact. {\bf 159} 351 (2004).
					
\bibitem{Kejun} Z. Kejun, J. Zhanpeng, J. Less-Common Met. {\bf 166}, 21 (1990).
					
\bibitem{Svechnikov}V. N. Svechnikov, A. K. Shurin and G. P. Dmitriyeva, Russ. Metall. Fuels {\bf 6}, 95 (1967).
					
\bibitem{Wodniecki}P. Wodniecki, B. Wodniecka, A. Kuli\'{n}ska, M. Uhrmacher, K.P. Lieb, J. Alloys Compd. {\bf 365}, 52 (2004).
					
\bibitem{PBlaha} P. Blaha, K. Schwarz, G.K.H. Madsen, D. Kvasnicka and J. Luitz, WIEN2k an Augmented Plane Wave Plus Local Orbitals Program for Calculating Crystal Properties, Vienna University of Technology, Vienna, Austria, 2001.
					
\bibitem{Perdew} J.P. Perdew, K. Burke and M. Ernzerhof, Generalized Gradient Approximation Made Simple, Phys. Rew. Lett. {\bf 77}, 3865 (1996).
					
\bibitem{Blochl} P.E. Blochl, O. Jepsen and O.K. Andersen, Phys. Rev. B {\bf 49}, 16223 (1994).
					
\bibitem{Kokalj} A. Kokalj, J. Mol. Graphics Modeling {\bf 17}, 176 (1999).
					
\bibitem{Murnaghan} F.D. Murnaghan, Proc. Natl. Acad. Sci. U.S.A. {\bf 30}, 244 (1944).
					
\bibitem{Herzig} P. Blaha, K. Schwarz and P. Herzig, Phys. Rev. Lett. {\bf 54}, 1192 (1985).            
					
\end{thebibliography}

\end{document}